\UseRawInputEncoding
\documentclass[fleqn,usenatbib]{mnras}
\usepackage[T1]{fontenc}
\usepackage{txfonts}
\usepackage{graphicx,times}
\usepackage{natbib}
\usepackage{amssymb}
\usepackage{longtable}
\usepackage{subfigure}
\usepackage{cite}
\usepackage{caption}

\title[Age-metallicity dependent stellar kinematics of the Milky Way disc from LAMOST and Gaia ]{Age-metallicity dependent stellar kinematics of the Milky Way disc from LAMOST and Gaia}
\author[Yaqian Wu]{Yaqian Wu$^{1,2}$\thanks{E-mail: wuyaqian@nao.cas.cn}, Maosheng Xiang$^{3,1}$\thanks{E-mail:mxiang@mpia-hd.mpg.de}, Yuqin Chen$^{1,4}$\thanks{E-mail:cyq@nao.cas.cn}, Gang Zhao$^{1,4}$, Shaolan Bi$^{2}$,
 \newauthor  Chengdong Li $^{5}$, Yaguang Li$^{6,7}$, Yang Huang$^{8,9}$ \\
$^{1}$CAS Key Laboratory of Optical Astronomy, National Astronomical Observatories, Chinese Academy of Sciences£¬Beijing 100101, P.\ R.\ China;\\
$^{2}$Department of Astronomy, Beijing Normal University,
             Beijing 100875, P.\ R.\ China;\\
$^{3}$Max-Planck Institute for Astronomy, K\"{o}nigstuhl 17, D-69117 Heidelberg, Germany;\\
$^{4}$School of Astronomy and Space Science, University of Chinese Academy of Sciences
             Beijing,  101408, P.\ R.\ China;\\
$^{5}$Rudolf Peierls Centre for Theoretical Physics, Clarendon Laboratory, Parks Road, Oxford, OX1 3PU, UK ;\\
$^{6}$Sydney Institute for Astronomy (SIfA), School of Physics, University of Sydney, NSW 2006, Australia ;\\
$^{7}$Stellar Astrophysics Centre, Department of Physics and Astronomy, Aarhus University, Ny Munkegade 120,
              DK-8000 Aarhus C, Denmark;\\
$^{8}$South West Institute For Astronomy Research, Yunnan University
             Kunming 650500, P.\ R.\ China;\\
$^{9}$Department of Astronomy, Peking University,
             Beijing 100871, P.\ R.\ China;\\
             }
\begin{document}


\pagerange{\pageref{firstpage}--\pageref{lastpage}} \pubyear{2017}

\maketitle

\label{firstpage}

\begin{abstract}
We investigate the stellar kinematics of the Galactic disc in 7 $<$ $R$ $<$ 13\,kpc using a sample of 118\,945 red giant branch (RGB) stars from LAMOST and Gaia. We characterize the median, dispersion and skewness of the distributions of the 3D stellar velocities, actions and orbital parameters across the age-metallicity and the disc $R$ -- $Z$ plane. Our results reveal abundant but clear stellar kinematic patterns and structures in the age -- metallicity and the disc $R$ -- $Z$ plane. The most prominent feature is the strong variations of the velocity, action, and orbital parameter distributions from the young, metal-rich thin disc to the old, metal-poor thick disc, a number of smaller-scale structures --- such as velocity streams, north-south asymmetries, and kinematic features of spiral arms --- are clearly revealed. Particularly, the skewness of $V_{\phi}$ and $J_{\phi}$ reveals a new substructure at $R\simeq12$\,kpc and $Z\simeq0$\,kpc, possibly related to dynamical effects of spiral arms in the outer disc. We further study the stellar migration through analysing the stellar orbital parameters and stellar birth radii. The results suggest that the thick disc stars near the solar radii and beyond are mostly migrated from the inner disc of $R\sim4 - 6$\,kpc due to their highly eccentrical orbits. Stellar migration due to dynamical processes with angular momentum transfer (churning) are prominent for both the old, metal-rich stars (outward migrators) and the young metal-poor stars (inward migrators). The spatial distribution in the $R$ -- $Z$ plane for the inward migrators born at a Galactocentric radius of $>$12\,kpc show clear age stratifications, possibly an evidence that these inward migrators are consequences of splashes triggered by merger events of satellite galaxies that have been lasted in the past few giga years.
\end{abstract}


\section{Introduction}

Understanding the formation and evolution of the Galactic disc is an interesting issue of contemporary astrophysics. Modern large-scale sky surveys, such as the Gaia astrometry survey \citep[e.g.][]{Brown16,Gaia16,Gaia18,Lindegren18}, and a number of Galactic spectrocopic surveys, such as the RAdial Velocity Experiment \citep[RAVE;][]{Steinmetz06}, SDSS \citep{Yanny09},
LAMOST \citep{Zhao06,Zhao12}, GALAH \citep{De15} and APOGEE \citep{Majewski17}, have revealed unprecedented details and phenomena of our Milky Way disc through the study of stellar distributions \citep[e.g.][]{Bovy12b,Widrow12,Mackereth17,Xiang18,Wang18a,Dobbie20}, chemistry \citep{Cheng12,Hayden14,Xiang15b,Kawata17,Frankel18,Ness18,Ness19,Wang19,Feuillet19,Lian20} and kinematics \citep{Lee11,Bovy12,Carlin13,Sun15,Aumer17b,Antoja18,Katz18,Kawata18,Tian18,Wang18b,Bland19,Katz19,Wh19,Yan19,Coronado20,Li20} of millions of stars. These studies have provided great insights on the structure, formation and chemo-dynamical evolution of the Galactic disc(s) \citep[e.g.][]{Minchev15,Aumer17,Ting19,Spitoni19}.

The Milky Way is suggested to have two major components, the thin and the thick disc, which differ in various ways, such as spatial distribution, age, metallicity and kinematics \citep[e.g.][]{Gilmore83,Bensby03,Spagna10,Lee11,Haywood13,Hayden14,Xiang15b,Xiang17a}. On kinematics, the thin disc contains stars that show larger rotation velocities and smaller velocity dispersion than the thick disc \citep[e.g.][]{Chiba00}.
 Stars in the thick disc at the solar neighbourhood are found to exhibit a positive relation between rotation velocity and metallicity, while a negative relation is found in thin disc stars \citep{Spagna10,Lee11,Adibekyan13,Recio14,Grieves18,Peng18,Wojno18}.
Using these differences in stellar kinematics, a star could be categorized into the thin or the thick disc population \citep[e.g.][]{Bensby03,Li17}.

The stellar velocities are found to exhibit significant variations in both the radial and the vertical directions of the Galactic disc. In the radial direction, \citet{Siebert11} detected a radial velocity gradient in the local disc with the RAVE survey data. \citet{Sun15} revealed a complex radial velocity pattern with significant radial variations in the 8$<$$R$$<$12\,kpc with the LAMOST spectroscopic survey of the Galactic anti-center \citep[LSS-GAC;][]{Liu14,Yuan15}. Utilizing the LAMOST data, \citet{Tian17} also revealed the radial velocity variations in the outer disc clearly. With the Gaia DR2, \citet{Katz18} revealed the complexity of the velocity field of the Galactic disc. They also observed streaming motions in $V_{R}$, $V_{\phi}$ and $V_{Z}$. With APOGEE red giant stars, \citet{Eilers20} found clear radial velocity patterns across the whole disc of 0$<$$R$$<$20\,kpc. These radial patterns were suggested to be related to the perturbation induced by both the rotating bar and the spiral arms.

In the vertical direction, the velocities of stars in the outer disc show strong north-south asymmetry \citep[e.g.][]{Carlin13,Sun15}. \citet{Williams13} and \citet{Sun15} investigated the stellar bulk motions of the nearby disc in detail, and revealed interesting patterns such as bending, breathing, and ripples. With LAMOST red clump stars, \citet{Wang18b} suggested that the outer disc exhibits complex kinematic structures. For example, there is a high radial velocity peak at $Z$ $\approx$ 0.5\,kpc, $R$ $\approx$ 10--11\,kpc. The complex structures are likely associated with over-density in the stellar number density \citep[e.g.][]{Xu15} and stellar mass distribution \citep[e.g.][]{Xiang18}. The outermost disc shows a strong warp structure \citep[e.g.][]{Skowron19, ChenXD19}, which exhibits significant features in stellar velocities \citep{Huang18, Poggio18, Romero19, Wang20}.

Radial migration is expected to significantly alter the position of stars in the Galactic disc \citep{Sellwood02,Roskar08,Loebman11,Minchev13,Grand16,A18}. Generally, there are two widely suggested processes responsible for the radial migration: one is the so-called blurring, which changes the stellar radial position due to epicycle motion without changing the angular momentum; the other is the so-called churning, which alters the Galactocentric position of the stars due to interactions with the Galactic bar or the spiral arms without significantly changing their orbital eccentricities \citep[e.g.][]{Schonrich09}. Thick disc stars may have more blurring than thin disc stars due to their highly eccentrical motions. Thin disc stars, however, may suffer strong churning effect. At the solar neighborhood, previous works have revealed the existence of super metal-rich stars (SMR), which have metallicities exceeding the present-day interstellar medium. These stars are likely a consequence of the radial migration due to the churning process \citep[e.g.][]{Nordstrom04,Boeche13, Kordopatis13, Xiang15a,Anders17,Chen19,Wang19}.

In this work, we present a study on stellar kinematics of the Galactic disc in 7 $<$ $R$ $<$ 13\,kpc with a large sample of red giant branch (RGB) stars from the LAMOST Galactic surveys. We characterize the median, dispersion and skewness of the distributions of the 3D stellar velocities, actions and orbital parameters across the age -- metallicity and the disc $R$ -- $Z$ plane. We further study the blurring and churning effects through analysing the stellar orbital parameters and stellar birth radii. The paper is organized as follows. Section\,2 introduces the data we adopted. Section\,3 presents the velocities and actions across the age -- metallicity and $R$ -- $Z$ planes. The orbital parameters and stellar migration are investigated in Section\,4, followed by conclusions in Section\,5.

\section{Data} \label{sec:style}

We make use of the LAMOST RGB stars from \citet{Wu19}, who estimated stellar ages and masses for 640\,986 RGB stars selected from the forth data release of the LAMOST Galactic surveys \citep{Deng12,Zhao12}. The RGB stars are selected with criteria of $T_{\rm eff}$ $<$ 5500\,K and log\ $g$ $<$ 3.8\,dex, and they are further distinguished from red clump stars using the g-mode period spacing ($\Delta$ $P$) derived from the LAMOST spectra with a data-driven method based on kernal principal component analysis (KPCA), taking the asterosesimic measurements from the $Kepler$ \citep{Vrard16} as the training set. The stellar ages and masses are also estimated with the KPCA method, using a sample of 5376 RGB stars with asteroseismic ages from \citet{Wu18} as the training set. The typical age uncertainty is 20 per cent for stars with the spectral signal-to-noise ratios (SNR) higher than 50 per pixel. Note that, the stellar ages for metal-poor stars are underestimated due to bias inherited from the training set. Here we opt to update the age estimates utilizing the same method as \citet{Wu19} but removing the metal-poor stars with artificially young ages from the training set. As a result, we find that, although not perfectly, this reduces the systematics of the age estimates by about 1 -- 2\,Gyr at the metal-poor end ([Fe/H]$\lesssim$ $-$0.6\,dex).

Stellar parameters of the RGB stars, including the radial velocity $V_{\rm r}$, effective temperature $T_{\rm eff}$, surface gravity log\ $g$, metallicity [Fe/H], absolute magnitudes ${\rm M}_V$ and ${\rm M}_{K_{\rm s}}$, alpha-element to iron abundance ratio [$\alpha$/Fe], C and N abundances [C/H] and [N/H], as well as extinction $E_{B-V}$ and spectroscopic distance, are estimated with the LAMOST Stellar Parameter Pipeline at Peking University \citep[LSP3;][]{Xiang15a,Xiang17a}, publicly available as the LAMOST value-added catalogue \citep{Xiang17b}\footnote{http://dr4.lamost.org/doc/vac}. Given a spectral SNR higher than 50, the stellar parameters yielded by LSP3 have a typical uncertainty of 4\,km/s in $V_{\rm r}$,100\,K for $T_{\rm eff}$, 0.1\,dex for log\ $g$, 0.3\,mag for $M_{V}$ and $M_{Ks}$, 0.1\,dex for [Fe/${\rm H}$], [C/H] and [N/H], and better than 0.05\,dex for [$\alpha$/Fe] \citep{Xiang17b}.

We cross-match the LAMOST RGB sample stars with Gaia DR2 for astrometric parameters. Furthermore, instead of using the spectroscopic distance in the LAMOST value-added catalogue, we make use of the Gaia distances from \citet{Bailer18} (here after BJ18). Given Gaia DR2 coordinates, proper motions, the BJ18 distances and the LAMOST radial velocities, we compute the Galactocentric 3D velocities ($V_{R}$, $V_{\phi}$, $V_{Z}$), actions ($J_{R}$, $J_{\phi}$, $J_{Z}$), defined as the integral of the canonical momentum along the orbit
\begin{equation}
J_i = 1/2\pi\oint{p_idx_i},
\end{equation}
where $p_{i}$ are the conjugate momenta, the
orbital eccentricity ($e$), epicentre ($R_{apo}$) and pericentre radii ($R_{peri}$), minimal and maximal vertical distances from the disc ($Z_{max}$ and $Z_{min}$), and guiding-centre radii ($R_{g}$) with the {\em Galpy} \citep{Bovy15}, using the MWPotential2014. We adopt a right-hand Cartesian coordinate to calculate the space motions of our sample stars in cylindrical coordinates. $V_{R}$ is positive for motion away from the Galactic Centre, $V_{\phi}$ is positive in the direction of Galactic rotation (note that we have changed the direction of $V_{\phi}$, in this reference frame a retrograde rotation is indicated by $V_{\phi}$ $<$ 0 km s$^{-1}$), and $V_{Z}$ is positive towards the north Galactic pole. The Sun is assumed to be located at (X, Y, Z)= ($-$8, 0, 0) kpc, and the solar motion w.r.t. the local standard of rest is ($U_{\odot}$, $V_{\odot}$, $W_{\odot}$) = (7.01, 10.13, 4.95) km s$^{-1}$ \citep{Huang15}. As a result, we obtained 568\,590 LAMOST-Gaia RGB stars with $X$, $Y$, $Z$, $V_{R}$, $V_{\phi}$, $V_{Z}$, $J_{R}$, $J_{\phi}$, $J_{Z}$, $e$, $Z_{max}$, $Z_{min}$, $R_{apo}$, $R_{peri}$, $R_{g}$.

\begin{figure*}
\centering
\includegraphics[width=\linewidth]{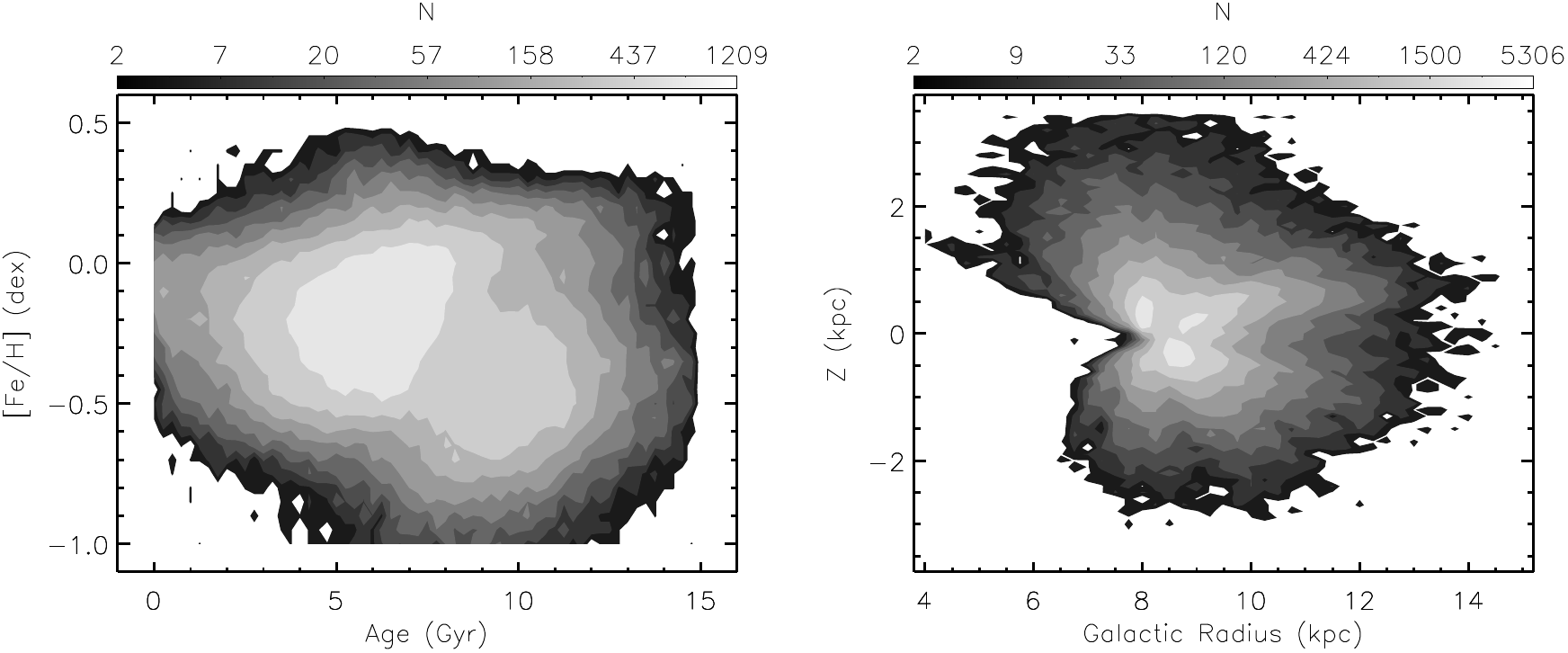}
\caption{Colour-coded stellar number density distributions of the LAMOST RGB sample stars in the age -- [Fe/H] ({\em left}) and the Galactic $R$ -- $Z$ ({\em right}) planes. The adopted bin size is 0.25\,Gyr $\times$ 0.05 dex for the age -- [Fe/H] plane, and 0.1\,kpc $\times$ 0.1\,kpc for the $R$ -- $Z$ plane.}
\label{Fig:1}
\end{figure*}
To obtain a clean sample of disc stars with high quality of age estimates, we further make a few quality cuts. First of all, to ensure good age estimates, we discard stars with low spectral quality, for those of spectral SNR $<$ 50 or stars with $d_{g}$ $<$0.8, where $d_{g}$ is an indicator of the similarity between the training and the target spectra defined in \citet{Wu19}. This cut removes 65 per cent stars in our sample.
Further we discard stars with log\,$g$ $<$ 2.0\,dex, as well as stars with [Fe/H] $<$$-$1\,dex, because in these parameter spaces, there is a lack of calibration stars with good seismic ages. In this work, we focus on a characterization of the observed kinematical properties of the disc stars. To minimize contamination from the halo stars, we remove stars with $V_{\phi}$ $<$ 50 km s$^{-1}$ from our sample. We further discard stars with parallax error larger than 20 per cent to ensure good quality of the kinematic data. Ultimately, our final sample contains 118\,945 stars. Fig.\,1 shows the finally adopted RGB stars in the age -- [Fe/H] space and the disc $R$ -- $Z$ plane. The left panel illustrates that the sample stars cover a wide age range, from 0 to $\sim$\,14\,Gyr. For all ages, the [Fe/H] exhibits a broad distribution. For stars older than 8\,Gyr, the bulk stars exhibit a negative trend between age and [Fe/H]. The right panel illustrates that the sample covers a Galactocentric radial distance of 5--14\,kpc, and a vertical distance of $-$3 to 3.5\,kpc from the disc mid-plane. About 84 per cent of the stars are located with 2\,kpc from the Sun.

Fig.\,2 shows the distribution of $R$ and $Z$ for our RGB stars in the age -- [Fe/H] plane. It shows that the metal-poor young stars have a median Galactocentric radius larger than 10\,kpc, while the other stars are mainly located at a Galactocentric radius not far from the solar value. In the vertical direction, as expected, the young, metal-rich thin disc stars are located close to the disc mid-plane, while the old, metal-poor thick disc stars are distributed at larger heights in general.

\begin{figure*}
\centering
\includegraphics[width=\linewidth]{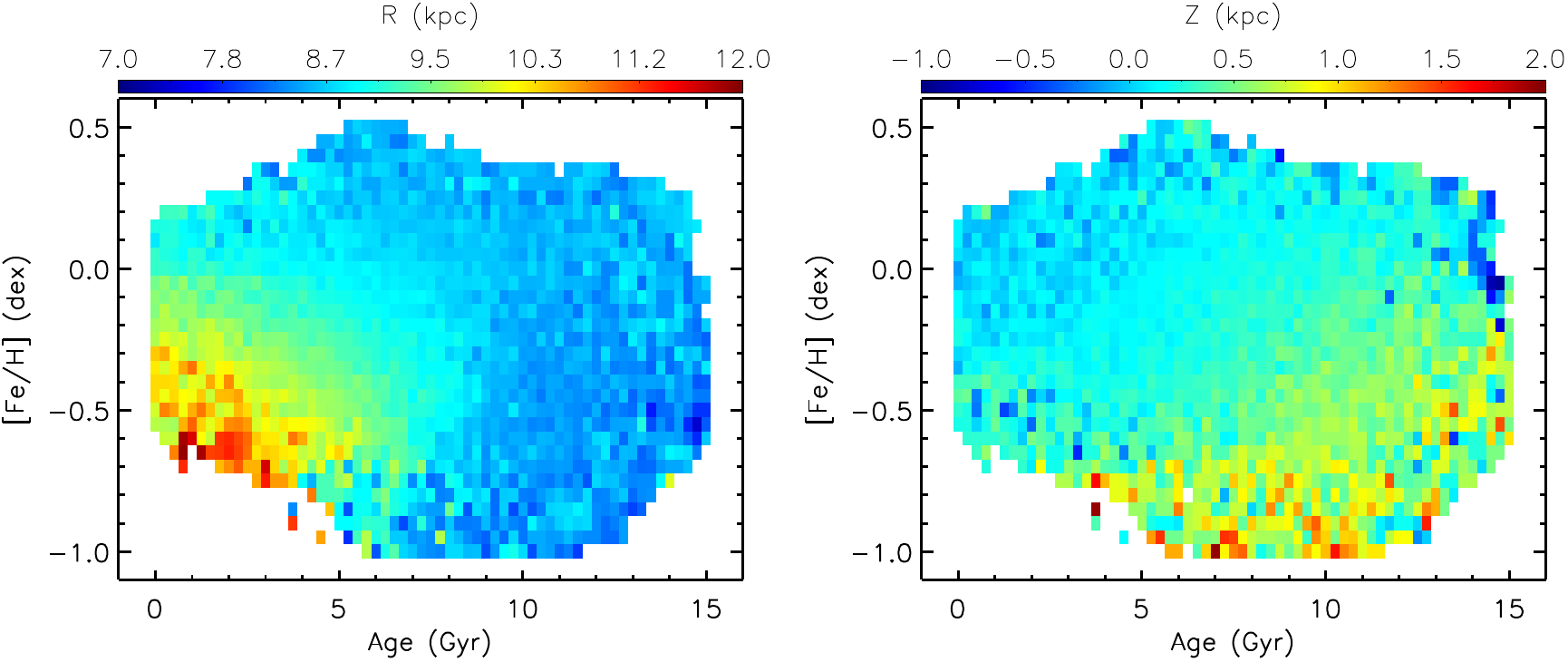}
\caption{[Fe/H] versus age diagram for the LAMOST-Gaia RGB stars, colour-coded by median values of $R$ and $Z$ for stars in each 0.25\,Gyr $\times$ 0.05 dex bin. }
\label{Fig:1}
\end{figure*}

\section{Velocities and actions}

In this section, we present the distributions of velocities and actions on the age -- metallicity plane in section 3.1 and on the $R$ -- $Z$ plane in section 3.2.
\subsection{Velocities and actions across the age -- metallicity plane}

 We investigate the distributions of $V_{R}$, $J_{R}$, $V_{\phi}$, $J_{\phi}$, $V_{Z}$ and $J_{Z}$ for mono-age and mono-metallicity stars. In steady state, the actions are invariant quantities describing the orbit of the star, while the velocities change along the orbit. Therefore a comparison of the distribution patterns between velocities and actions for mono-age and mono-metallicity stars tells us information about the impact of stellar motion on the morphology of stellar distribution in the age--metallicity plane. A comparison of the two also reflects the impact of dynamical evolution due to angular momentum transfer, as the resultant alteration may differ between the velocity distribution and action distribution. For example, $'$churning$'$ process may significantly increase the angular momentum, but may not significantly alter the rotation. difference caused by orbital motion.

 We divide the age-[Fe/H] plane into individual bins, and characterize the distribution of the velocities and actions in each individual bin using three parameters -- median, dispersion and skewness. Here the dispersion is defined as the standard deviation, and the skewness is defined as
 \begin{equation}\label{1}
  \alpha_{i}=(Ei^3-3\mu\sigma^2-\mu^3)/\sigma^3,
\end{equation}
 where $\mu$ represents the median and $\sigma$ represents the standard deviation.
To calculate the median, dispersion and skewness of the distribution, we adopt a bin size of 1\,Gyr $\times$ 0.1\,dex. We require the minimum number of stars in each bin to be at least 100. Bins with fewer stars are neglected.

\subsubsection{the radial kinematics}
Fig.\,3 plots the results for $V_{R}$ and $J_{R}$. It shows that the median values of $V_{R}$ in individual bins are small ($-10\lesssim V_{R} \lesssim10$\,km/s) across the whole age -- [Fe/H] plane. There are no strong variations across the plane except for the young, metal-poor stars, which exhibit larger mean radial motion than the other stars. These young, metal-poor stars are found to be thin disc stars in the outer disc, and they also exhibit strong stream motions in the disc $R$ -- $Z$ plane (Fig.\,6). The mean $J_{R}$ shows a clear separation in the age -- [Fe/H] plane: the younger, more metal-rich stars exhibit small $J_{R}$, while the older, metal-poor stars exhibit large $J_{R}$. This suggests that the younger, more metal-rich stars indeed have small radial motion. While the old, metal-poor stars have strong radial motion, and their small mean $V_{R}$ value is simply due to an average effect. Although the young, metal-poor stars have relatively large mean $V_{R}$, they have small $J_{R}$, which may be an evidence that these stars have experienced perturbations.
The sharp border is a good distinction between the thin and thick disc populations --- the more metal-poor, older stars with larger $J_{R}$ are dominated by the thick disc stars, which have larger orbit eccentricities than the more metal-rich younger thin disc stars, which are dominated by circular orbits (thus small $J_{R}$), \citep[e.g.][]{Jing16, Mackereth17, Feuillet19, Yan19}. This can also be seen in Fig.\,6, in which the stars with large $J_{R}$ are distributed at large heights above the disc plane. For convenience, hereafter, we will use the terms `thin disc' and `thick disc' to qualitatively represent the younger, more metal-rich and older, more metal-poor stars, respectively, as separated by the border. Note that the old ($\gtrsim10$\,Gyr), metal-rich (${\rm [Fe/H]}\gtrsim0$) stars also exhibit small $V_{R}$ and $J_{R}$ values, implying that these thin disc stars are co-formed the metal-poor thick disc stars since they have comparably old ages. Finally, we note that the stellar age for thick disc stars at the metal-poor end might have been underestimated due to a lack of seismic calibration stars. While this does not affect the main conclusions in the current paper as the difference of kinematics between the thin and thick disc stars are obvious.

The dispersion of both $V_{R}$ and $J_{R}$ exhibit clear increasing trend from the young metal-rich stars to the old, metal-poor stars. For populations with $\tau\lesssim8$\,Gyr and ${\rm [Fe/H]}>0$\,dex, which are mostly composed of the $'$thin$'$ disc stars, the dispersion is mainly a function of stellar age, while the trend with [Fe/H] is not obvious. Since there are strong radial metallicity gradients for these relatively young stellar populations \citep[e.g.][]{Xiang15b,Anders17}, a lack of trend with [Fe/H] implies that the dispersion for mono-age and mono-metallicity stellar population does not change significantly with Galactocentric radius. This conclusion can be also reached by combining Fig.\,3 with Fig.\,2, which shows that the stars with age $<$ 5\,Gyr and [Fe/H] $<$ $-$0.2\,dex have a mean Galactocentric radius beyond 10\,kpc, while the stars with solar metallicity at the same age mainly locate at $\sim$\,8\,kpc. Such a lack of radial variations in $V_{R}$ and $J_{R}$ is probably related to the bifurcation structure in the $V_{R}$ distribution of the outer-disc stars\citep{Liu12,Sun15}, as these bifurcation structures will increase the measured dispersions at the outer disc thus the metal-poor bins of a given age.

 The skewness of $V_{R}$ is $\sim$ 0 and it uniformly distributes across the age -- [Fe/H] plane, indicating that for all mono-age and mono-abundance bins, the $V_{R}$ follows nearly Gaussian distributions (See Appendix, Fig.\,A1). However, the skewness in $J_{R}$ shows strong patterns, as the young, metal-rich stars show much larger positive skewness than the old, metal-poor stars. Note that given the definition of the $J_{R}$, the skewness is always positive (Fig.\,A1). Interestingly, Fig.\,3 shows that the skewness of the old, metal-rich stars are different from the young metal-rich stars but similar to the old, metal-poor stars, in contrast to the case of mean $J_{R}$. We find that the $J_{R}$ distribution for the old, metal-rich stars is more dispersed than the young, metal-rich stars. It makes the skewness of the old, metal-rich stars lower than the young metal-rich stars.

\begin{figure*}
\centering
\includegraphics[width=\linewidth]{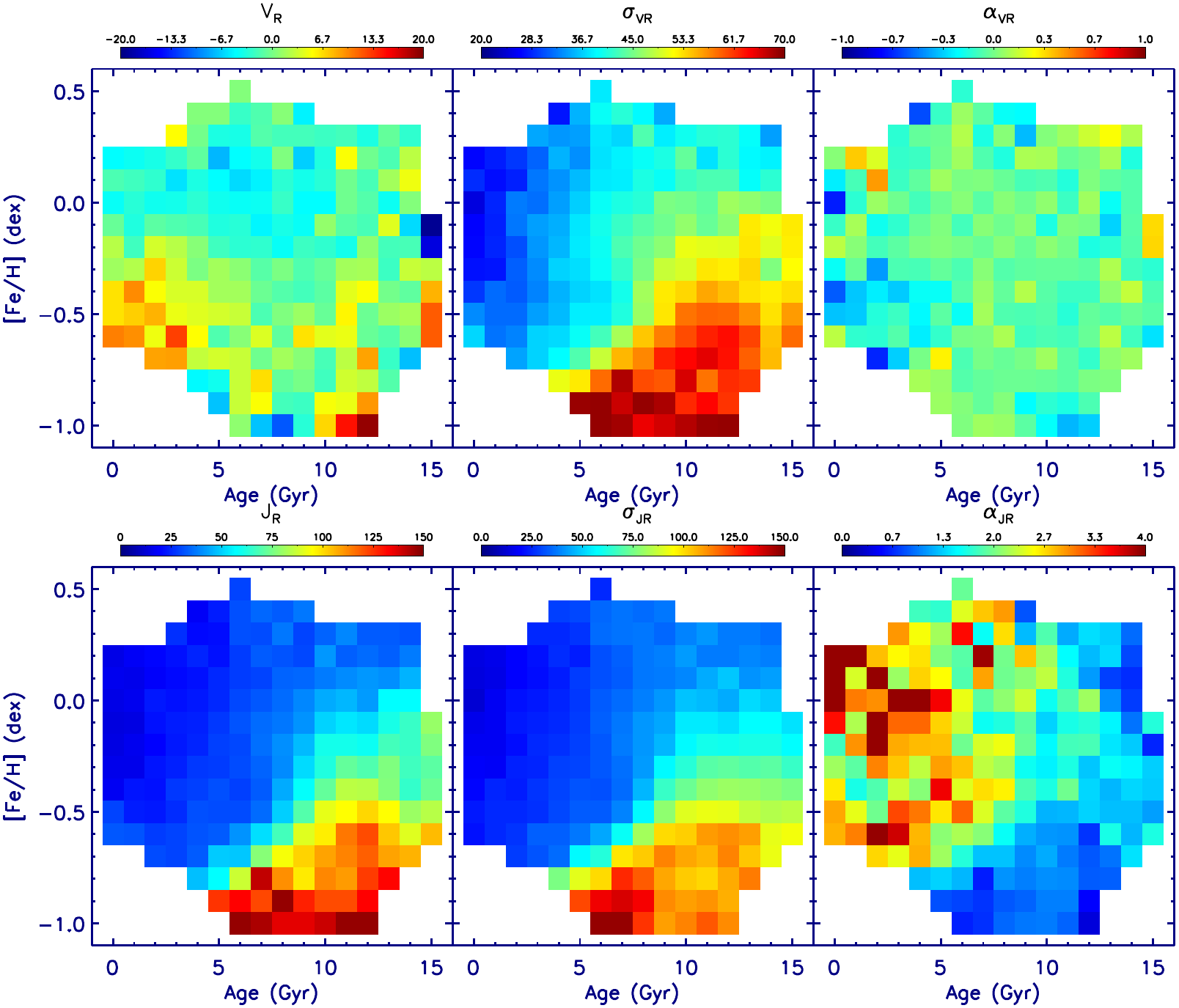}
\caption{Distributions of the LAMOST-Gaia RGB stars in the age -- [Fe/H] plane, colour-coded by the values of median ({\em left}), dispersion ({\em middle}), and skewness ({\em right}) of the distribution functions of the radial velocity ($V_{R}$) and action ($J_{R}$) for stars in each 1.0\,Gyr $\times$ 0.1\,dex age -- [Fe/H] bin. \label{fig:fig1}}
\end{figure*}

\subsubsection{azimuthal kinematics}
Fig.\,4 plots the results of $V_{\phi}$ and $J_{\phi}$.
It shows that the median $V_{\phi}$ and $J_{\phi}$ are clearly different between the younger, more metal-rich thin disc stars and the older, more metal-poor thick disc stars. As the latter has lower rotation velocities and angular momentum. The sharp border between the two overall populations (i.e., the thin disc and the thick disc) is consistent with Fig.\,3. For the thin disc stars, the values of $V_{\phi}$ and $J_{\phi}$ increase with decreasing [Fe/H]. In contrast, the values for $V_{\phi}$ and $J_{\phi}$ of the thick disc stars exhibit a positive trend with [Fe/H]. This is consistent with previous results \citep{Spagna10, Lee11, Adibekyan13, Recio14, Allende16,Peng18,Yan19}. The figure shows such a negative trend for thin disc stars of all mono-age stellar populations of $\tau\lesssim7$\,Gyr. The reason of such a negative trend between $V_{\phi}$ and [Fe/H] is likely a combined consequence of the stellar migration due to epicycle motion and the negative radial metallicity gradients of the Galactic disc. That is to say, the outer, metal-poor stars are mainly inward migrators, at their peri-center phase of their orbits (see Section 4) and with large $V_{\phi}$, while the inner/solar neighbourhood, metal-rich stars are mainly outward migrators, at their apo-center phase of their orbits and with small $V_{\phi}$ \citep[see e.g.][]{Loebman11, Curir12, Minchev15, Kawata18}. For the thick disc stars, the positive trend between $V_{\phi}$ and [Fe/H] is likely a consequence of a negative vertical metallicity gradient \citep[e.g.][]{Curir12,Xiang15b,Anders17,Kawata18} and the stellar migration. The more metal-poor thick disc stars are generally at larger heights and have smaller $V_{\phi}$ either because they were born hotter so that they have larger eccentricity and/or because the thick disc was born with a positive radial metallicity gradient \citep[e.g.][]{Curir04}.
For a given [Fe/H], younger stars have generally larger $V_{\phi}$ or $J_{\phi}$, which is consistent with the fact that older stars exhibit more dynamical evolution \citep[e.g.][]{Ting19}. Note that at the metal-poor and $\tau<5$ Gyr region, the data present a sharp boundary that exhibits a negative trend between [Fe/H] and age. This is likely a consequence of Galactic chemical enrichment with time \citep[e.g.][]{Schonrich09}. This effect also causes a steeper $V_{\phi}$ -- [Fe/H] gradient for the younger population.

The dispersions of $V_{\phi}$ and $J_{\phi}$ also exhibit clear increasing trend from the younger, more metal-rich stars to the older, more metal-poor stars. For the thin disc population, the dispersion of $V_{\phi}$ does not change significantly with [Fe/H], indicating that the values for $\sigma_{V_{\phi}}$ is nearly constant along the Galactocentric radius for mono-age stellar population. However the values of $\sigma_{J_{\phi}}$ at a given age increases with decreasing [Fe/H]. This is simply a consequence of constant $\sigma_{V_{\phi}}$ and the fact that the more metal-poor stars are located at larger Galactocentric radius.

The skewness of $V_{\phi}$ is mostly negative for thin disc stars and $\sim$\,0 for thick disc stars. This means that the $V_{\phi}$ distribution for the thin disc stars has a negative tail, which is largely contributed by stars in the outer disc (i.e., with larger $R$). While the $V_{\phi}$ distribution for the thick disc stars is nearly Gaussian. The skewness of $J_{\phi}$ is close to zero in the majority of the age--[Fe/H] bins, even for many of which the $V_{\phi}$ skewness is negative. This is not surprising, because some of stars in the negative tail of $V_{\phi}$ are located at larger $R$, so that the $J_{\phi}$ ($=V_{\phi}$ $\times$ $R$) distribution will not exhibit a similar negative tail to the $V_{\phi}$. Nonetheless, for some age -- [Fe/H] bins of $\tau$ $<$ 5\,Gyr, ${\rm [Fe/H]}\sim-0.6$\,dex, the stars exhibit strong negative skewness in $J_{\phi}$, possibly due to spiral arm dynamics (Wu et al. in preparation, see also Kawata et al. 2018).

\begin{figure*}
\centering
\includegraphics[width=\linewidth]{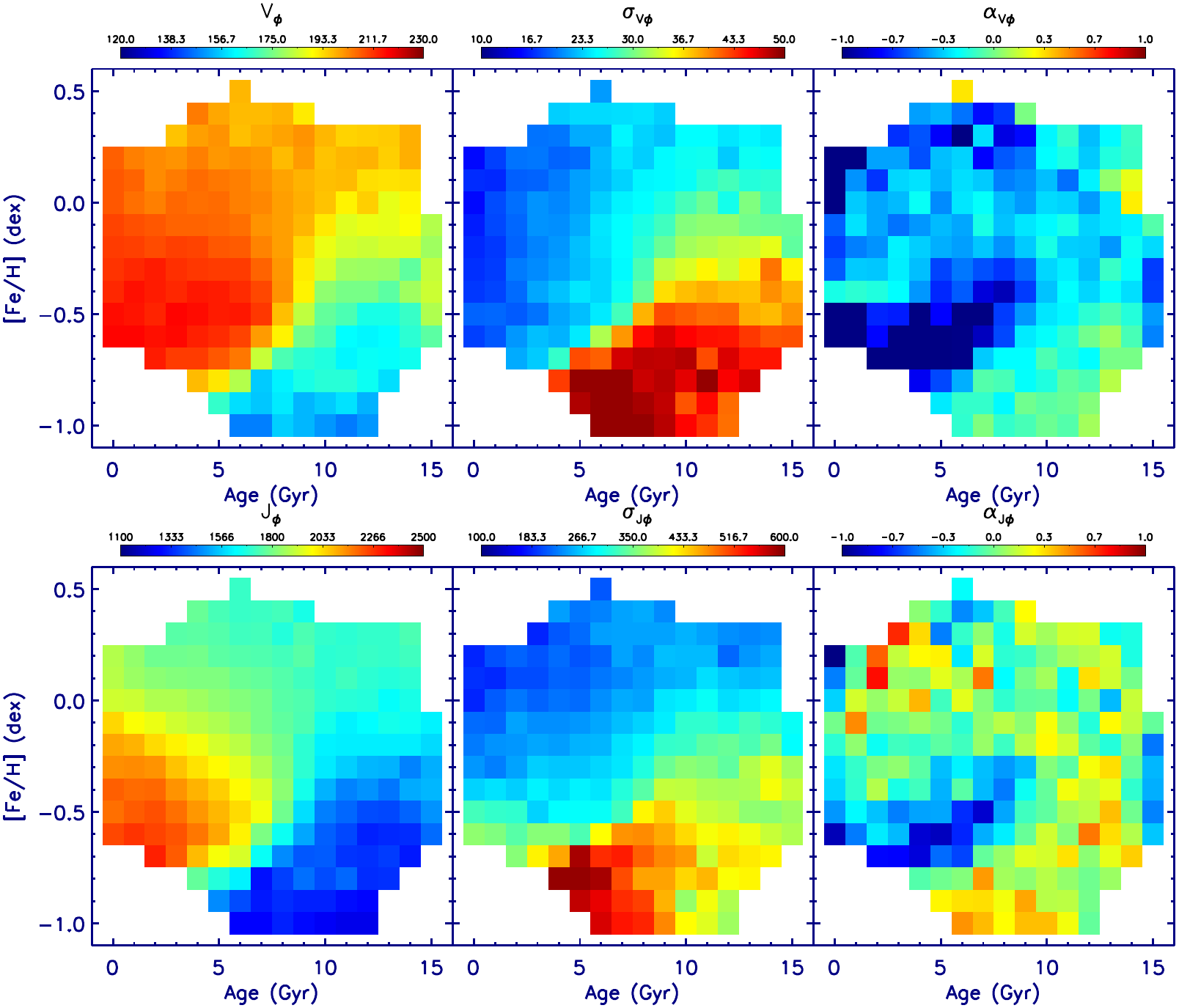}
\caption{Same as Fig.\,2, but for the azimuthal component of the stellar velocities and actions. \label{fig:fig1}}
\end{figure*}

\subsubsection{vertical kinematics}
Fig.\,5 shows that the median value of $V_{Z}$ is generally small ($-$5\,km/s\,$\lesssim V_{Z}\, \lesssim5$\,km/s), and largely across the whole age -- [Fe/H] plane expect for the young, metal-poor stars, which exhibit systematically positive mean motion. The $J_{Z}$ exhibits a clear increasing trend from the younger, metal-rich thin disc stars to the older, metal-poor thick disc stars, and there is a clear border between the thin and thick stars.

The dispersion of $V_{Z}$ increases smoothly with both age and [Fe/H]. For stars with the same age, the velocity dispersion increases as [Fe/H] decreases. At a given [Fe/H], the dispersion of $V_{Z}$ increases with age for the `thin' disc stars. This relation is known as the age -- velocity dispersion relation (AVR), the results are consistent with previous works \citep[e.g.][]{Holmberg09,Tian15,Hayden17,Hayden20}. The observed AVR of the thin disc stars suggested to be a consequence of gradual heating through scattering process, and the stars were born with roughly constant velocity dispersion of smaller than 10\,km s$^{-1}$ \citep{Aumer09,Ting19}.
Similar to the case of mean $J_{Z}$, the dispersion of $J_{Z}$ shows a sharp border between the younger, more metal-rich stars and the older, more metal-poor stars.

Similar to the case of mean $V_{Z}$, the skewness of $V_{Z}$ is also $\sim$ 0 and uniformly distributes across the age -- [Fe/H] plane, indicating that for all mono-age and mono-abundance bins, the $V_{Z}$ follows nearly Gaussian distributions (Fig.\,A1). While the skewness of $J_{Z}$ shows strong patterns, as the young metal-rich populations show larger positive skewness than the old, metal-poor populations. Note that given the definition of $J_{Z}$, the skewness is also always positive (Fig.\,A1).

\begin{figure*}
\centering
\includegraphics[width=\linewidth]{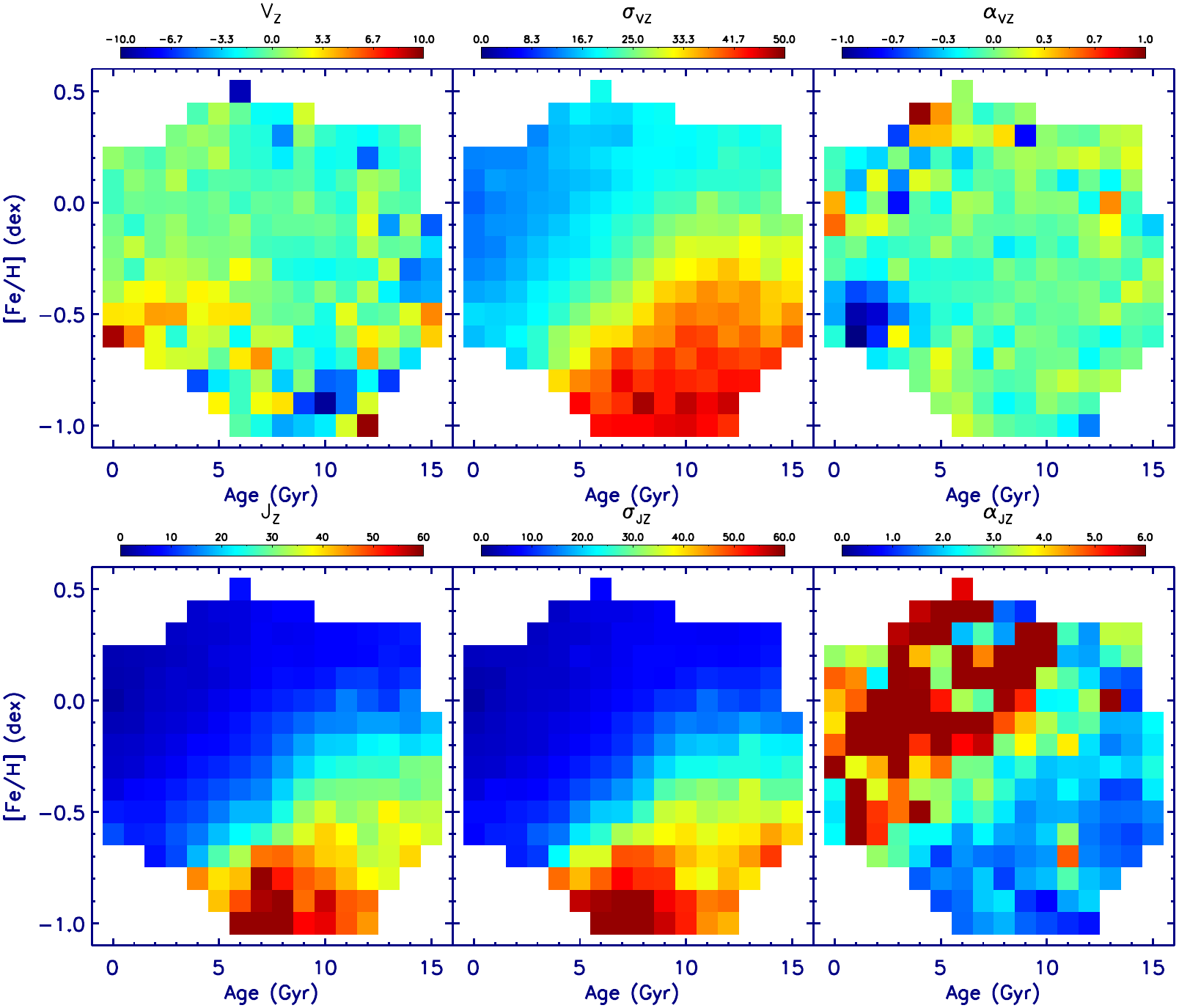}
\caption{Same as Fig.\,2, but for the vertical component of the stellar velocities and actions.  \label{fig:fig1}}
\end{figure*}
\subsection{Kinematics across the R -- Z plane}
\subsubsection{Radial velocity $V_{R}$ \& action $J_{R}$}
Fig.\,6 plots the results of $V_{R}$ and $J_{R}$ across the $R$ -- $Z$ plane. To calculate the median, dispersion and skewness of the distribution, we adopt a bin size of 0.5 $\times$ 0.5\,kpc. We further require a minimal number of 100 in each bin, and discard bins with fewer stars. The mean $V_{R}$ shows irregular substructures. At the solar radius of 7\,kpc $<$ $R$ $<$ 9\,kpc, and $|Z|$ $<$ 0.5\,kpc, stars with $R$ $<$ 8\,kpc move outwards (with positive velocity), while stars with $R$ $>$ 8\,kpc move inwards (with negative velocity). It shows a negative radial velocity gradients in the solar radius. Utilizing the RAVE data, \citet{Siebert11} measured a negative radial velocity gradient from about 2\,kpc inside of the Sun to about 1\,kpc outside. \citet{Williams13} confirmed the negative radial velocity gradients using RAVE red-clump stars. With LAMOST red-clump stars, \citet{Tian17} and \citet{Wang18a} found that the radial velocity is oscillating along $R$, as the value is negative at $R$ $<$ 9\,kpc but becomes positive beyond $R$ $\sim$ 9\,kpc. These patterns also revealed by \citet{Katz18}.
In the outer disc, the most prominent features are those with positive $V_{R}$ in $10<R<14$\,kpc.
 There are two positive velocity peaks, one at $10\lesssim R\lesssim12$\,kpc and $Z\sim0.5$\,kpc, the other at $R\sim13$\,kpc and $Z\sim0$\,kpc. These outer-disc substructures correspond to those presented in Fig.\,3 for the young, metal-poor stars. In Fig.\,7, we slice the stars into age bins to show how these $V_{R}$ structures vary with age. The figure illustrates that the substructure with positive $V_{R}$ at the outer disc of $Z\sim0.5$\,kpc is strong for young stellar populations, and is also present in old populations. This structure
is likely related to the so-called `north-near' overdensity/stream presented in previous works \citep{Xu15,Wang18a,Xiang18}. The peak at $R\sim13$\,kpc and $Z\sim0$\,kpc occurs in the age bin of 3-6\,Gyr, but it is hard to tell if they are also present in other age bins due to the limited spatial coverage of the data (but see Wang et al. 2018 for velocity structure in the outermost disc of $R>12$\,kpc).
The $J_{R}$ distribution exhibits a flaring structure in the outer disc, and the $J_{R}$ values are small near the Galactic plane across the whole radii. In contrast, the stars with $R<10$\,kpc and $|Z|\gtrsim1$\,kpc exhibit large $J_{R}$ values, suggesting that stars at those high Galactic latitude, mostly thick disc stars, have strong radial motions.

The dispersion of $V_{R}$ and $J_{R}$ exhibits a strong flaring structure across the whole radial range from $R$ $\sim$ 6\,kpc to $R$ $\sim$ 13\,kpc, with larger dispersion at larger heights. At $|Z|$ $<$ 1\,kpc, the dispersion of $V_{R}$ decreases significantly with increasing Galactic radius, which is likely driven by the age of the stellar populations --- the outer disc is dominated by younger stars.

 The skewness of $V_{R}$ is close to 0 across the $R$ -- $Z$ plane, except for the outer disc region of $R$$\gtrsim$11\,kpc, which shows negative skewness, possibly caused by the strong substructures with positive $V_{R}$. The $J_{R}$ for stars at low Galactic heights or outer disc, mostly thin disc stars, exhibits strong positive skewness due to the sharp distributions (see also Fig.\,3 and the according explanations in section 3.1.1).

\begin{figure*}
\centering
\includegraphics[width=\linewidth]{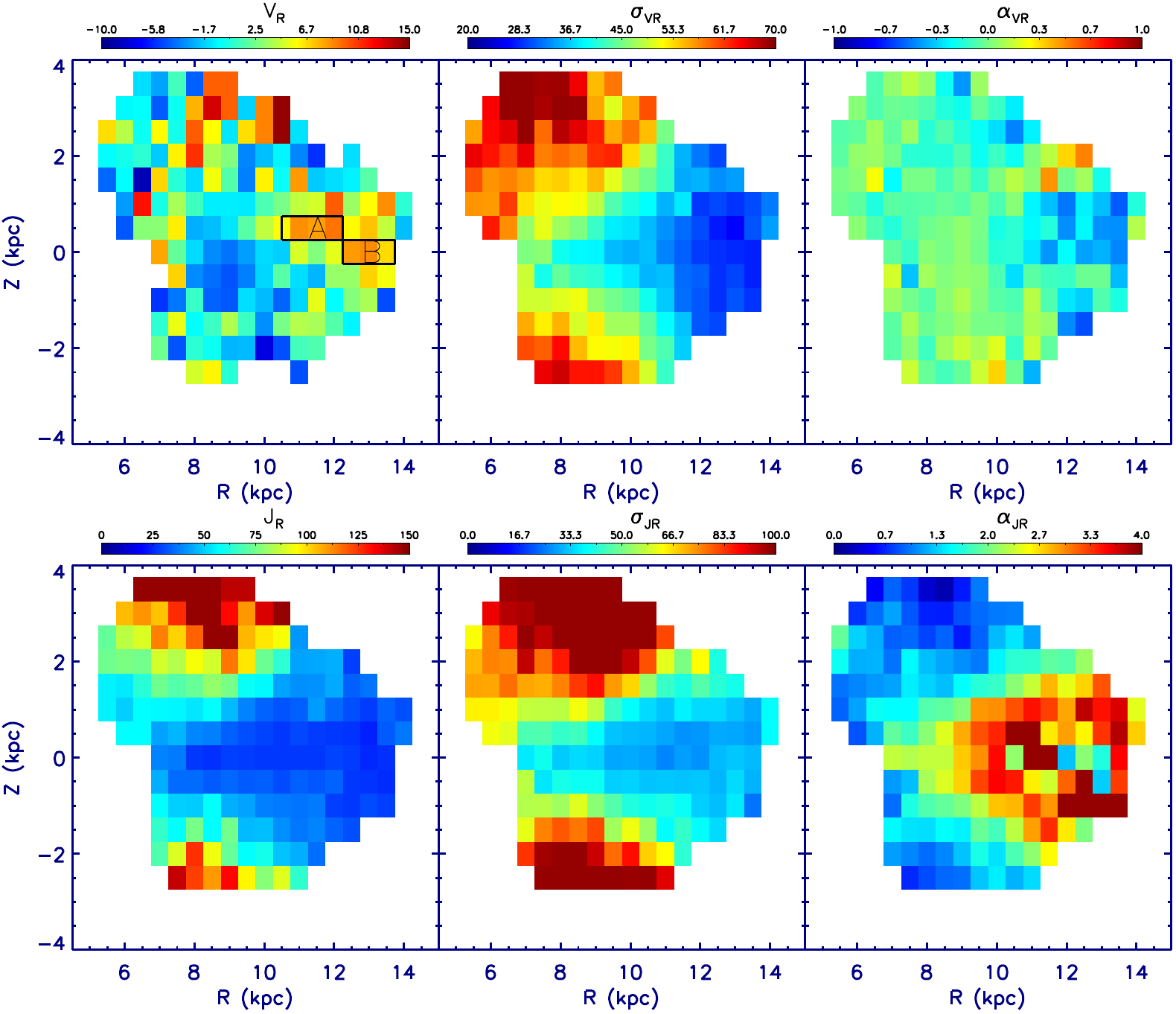}
\caption{Distributions of the LAMOST RGB stars in the Galactic $R$--$Z$ plane. Colours represent the median ({\em left}), dispersion ({\em middle}), and skewness ({\em right}) of the radial velocities and action distribution functions for stars in each 0.5\,kpc $\times$ 0.5\,kpc bin. The substructures (A \& B) with high $V_{R}$ values are marked in the top left panel (see text). \label{fig:fig1}}
\end{figure*}

\begin{figure*}
\centering
\includegraphics[width=\linewidth]{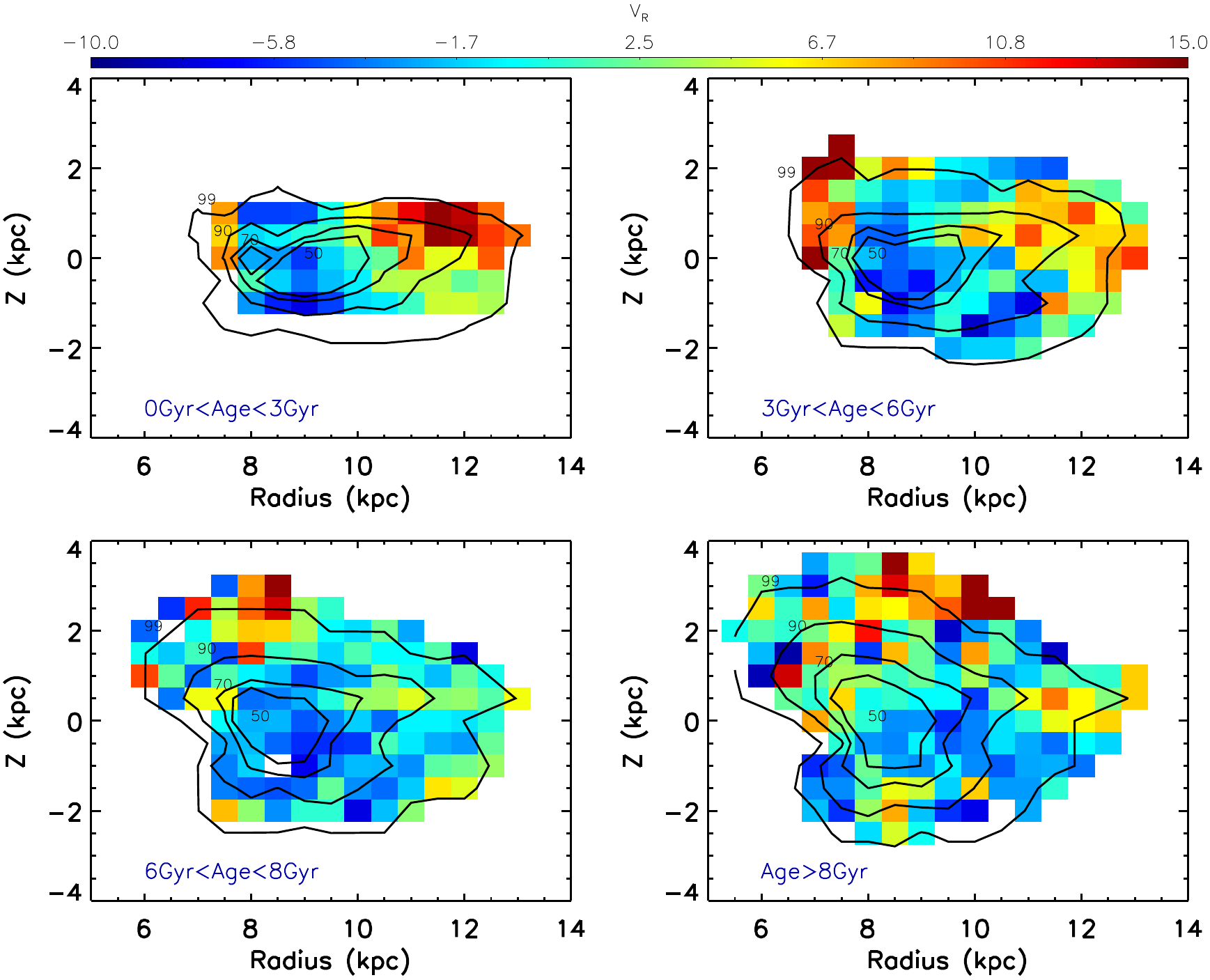}
\caption{Colour-coded distributions of median $V_{R}$ across the distance $R$--$Z$ plane for stars in different age bins. Also shown are the contours indicating 50,70,90, and 99 per cent of the sample stars.}
\end{figure*}

\subsubsection{Azimuthal velocity $V_{\phi}$ \& action $J_{\phi}$}
Fig.\,8 plots the results of $V_{\phi}$ and $J_{\phi}$ across the $R$ -- $Z$ plane.
It shows a clear vertical gradient on both sides of the disc that represents the transition from thin to thick disc. The $V_{\phi}$ of the thin disc exhibits a flaring structure.
In the outer disc of $R$ $\geq$ 10\,kpc and $|Z|$ $\geq$ 0.5\,kpc, stars in the south are rotating faster than those at the symmetric positions in the north, which is consistent with previous findings of \citet{Wang18a}.
The $J_{\phi}$ shows regularized flaring patterns across the whole disc covered by our data, a consequence of the increase of Galactocentric radius and the vertical $V_{\phi}$ gradient, as the former causes larger $J_{\phi}$ at outer disc while the latter causes smaller $J_{\phi}$ at larger height.

The dispersion of $V_{\phi}$ and $J_{\phi}$ also exhibits a clear increasing trend from the smaller heights to larger heights. For the thin disc stars, the dispersion of $V_{\phi}$ exhibits a clear flaring structure --- stars at larger $|Z|$ or smaller $R$ have larger $V_{\phi}$ dispersion. As a consequence, the dispersion of $J_{\phi}$ is largely invariant with $R$. However, for the thick disc stars, the dispersion of $V_{\phi}$ is largely invariant with $R$. As a consequence, the dispersion of $J_{\phi}$ increases significantly with $R$ in the thick disc.

The skewness of $V_{\phi}$ and $J_{\phi}$ is generally negative for thin disc stars and $\sim$ 0 for thick disc stars.
Interestingly, in 11 $<$ $R$ $<$ 13\,kpc, $-$0.25 $\lesssim$ $Z$ $\lesssim$ 0.25\,kpc, the skewness of $V_{\phi}$ and $J_{\phi}$ is clearly different from the other regions nearby. We find that there are bifurcations in the $J_{\phi}$ distribution that is probably an effect of spiral arms. The detailed discussions on this structure will be presented in Wu et al. (in preparation).
\begin{figure*}
\centering
\includegraphics[width=\linewidth]{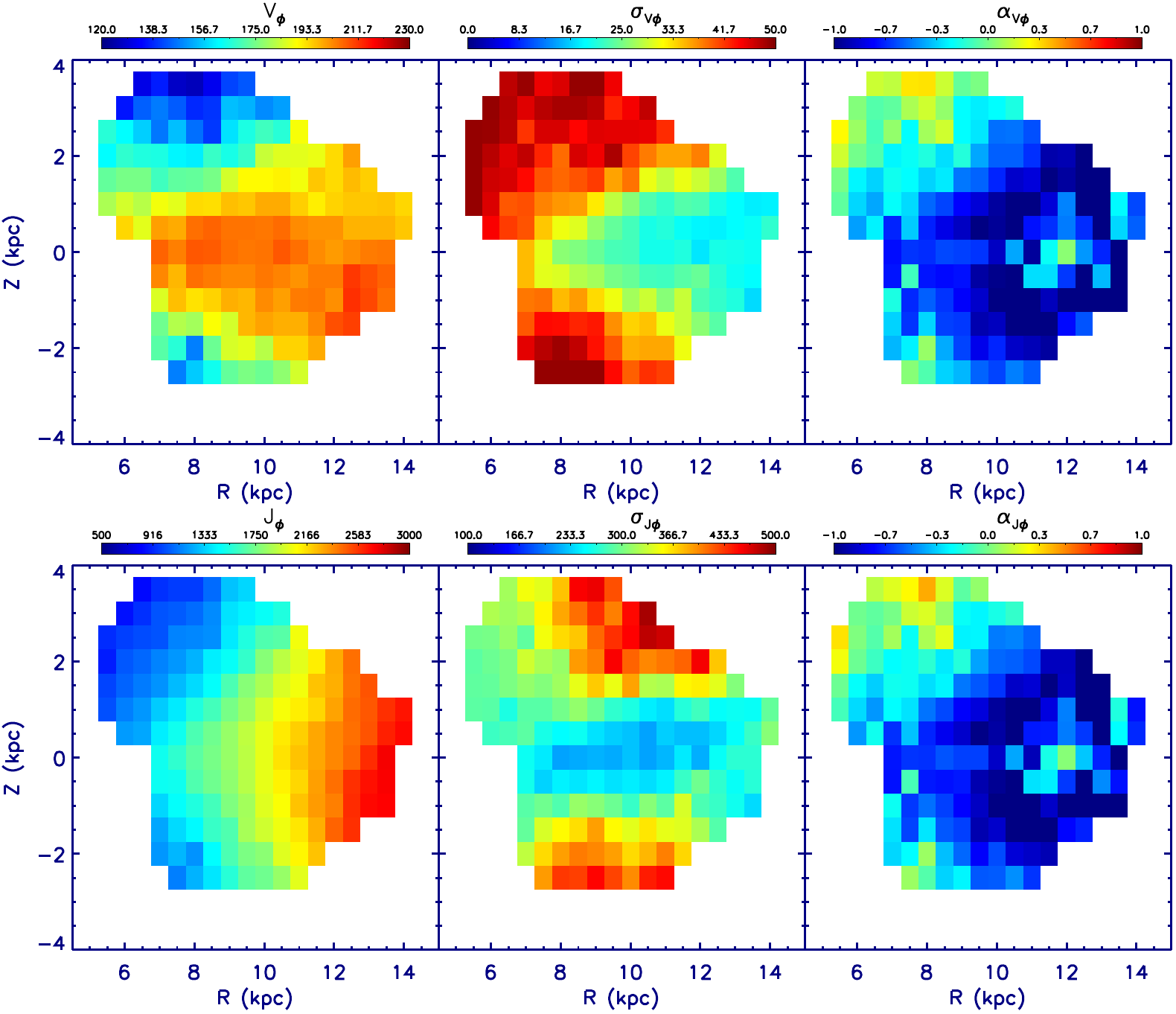}
\caption{Same as Fig.\,6, but for the azimuthal component of the stellar velocities and actions. \label{fig:fig1}}
\end{figure*}

\subsubsection{Vertical velocity $V_{Z}$ \& Action $J_{Z}$}
Fig.\,9 plots the results of $V_{Z}$ and $J_{Z}$ across the $R$ -- $Z$ plane. It shows that stars in the north and south sides of the plane at $R$ $\sim$ 8.5\,kpc exhibit net upward vertical motions. This is bending mode \citep{Widrow12,Williams13,Widrow14,Sun15,Carrillo18,Chequers18}, and the results are similar to the results shown by \citet{Williams13,Carrillo18,Wh19}.
At the outer disc $R$ $\sim$ 13\,kpc, It is revealed that the vertical velocities in southern disc stars have larger vertical velocities than the northern disc stars, due to the effect of the warp structure \citep{Skowron19,Chen19}. The values for $J_{Z}$ show a strong dependence on $Z$ while they are only marginally dependent on $R$. This is determined by the nature of $J_{Z}$. As $J_{Z}$ quantifies the vertical excursions of an orbit, and vanishes for orbits lying entirely in the Galactic plane, stars at large heights have large vertical excursions, while stars near the plane have large probabilities to be with small vertical excursions.

The dispersion of $V_{Z}$ exhibits a clear flaring structure, which shows a negative radial gradient and a positive vertical gradient. The negative radial gradient of $\sigma_{V_{Z}}$ is expected since $\sigma_{V_{Z}}$ associated with the surface mass density \citep[e.g.][]{Freeman02,Binney08},
which decreases with increasing Galactocentric radius \citep[e.g.][]{Xiang18}. The positive vertical gradient mainly reflects the increase of surface mass density enclosed by increasing height above the disc. The dispersion of $J_{Z}$, however, has less dependence with $R$. Again, this is likely due to the nature of $J_{Z}$, as discussed above.

The skewness of $V_{Z}$ is close to 0 and uniformly distributes across the $R$ -- $Z$ plane, except for the outermost region ($R>$11kpc), whereas the skewness in $J_{Z}$ shows strong patterns. These characteristics are similar to the case of skewness for $V_{R}$ and $J_{R}$, and they share similar physical reasons (see Section 3.2.1).
\begin{figure*}
\centering
\includegraphics[width=\linewidth]{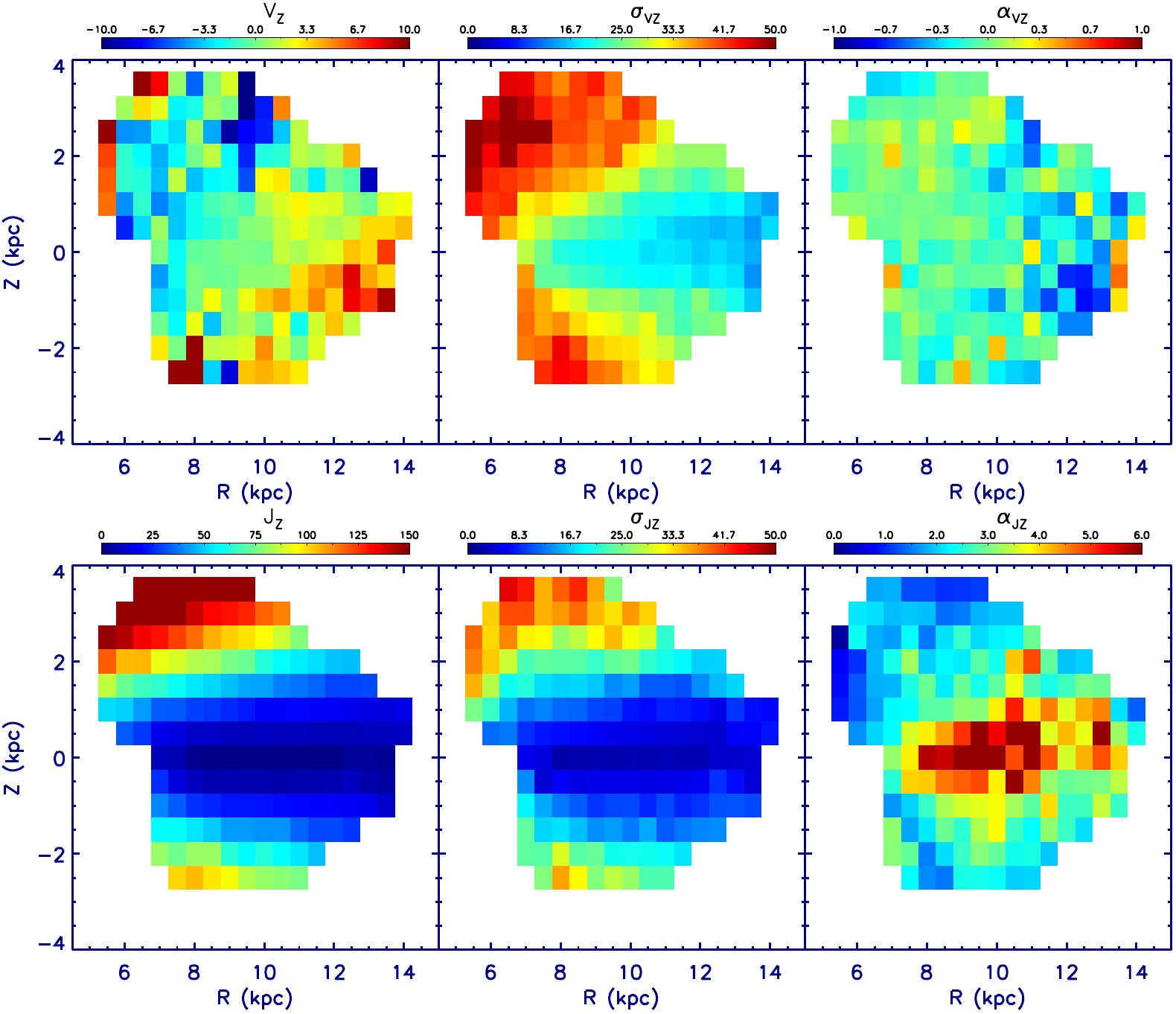}
\caption{Same as Fig.\,6, but for the vertical component of the stellar velocities and actions.  \label{fig:fig1}}
\end{figure*}

\section{Orbital parameters and stellar migration}

In this section, we present the orbital parameters of the RGB stars across the age -- metallicity plane and the disc $R$ -- $Z$ plane, and discuss the impact of non-circular orbital motion on the stellar distribution by comparing the current position of the stars with the guiding-centre radii of their orbits. We further estimate the birth radii of the stars based on their age and metallicity in the same way as \citet{Chen19}, initially proposed by \citet{Minchev18}, and study the effect of stellar migration (churning). Note that here we refer to the stellar migration as the churning process, in which stars change their orbital radii through angular momentum transfer via scattering or perturbation \citep[e.g.][]{Sellwood02,Minchev13}.

\subsection{Orbital parameters}

As mentioned in Section\,2, the orbital parameters of our sample stars are derived using positions and proper motions from Gaia DR2, distances from BJ18, radial velocities from LAMOST DR4. The orbit calculation is implemented with the {\em Galpy}, assuming the MWpotential2014 \citep{Bovy15}.

Fig.\,10 plots the median, dispersion, and skewness of the stellar eccentricity distributions in the age -- [Fe/H] plane and the disc $R$ -- $Z$ plane. The figure illustrates that the young or old but metal-rich thin disc stars have mean orbital eccentricities smaller than 0.2, and have small ($<$0.1) dispersion among stars in each mono-age and mono-[Fe/H] bin, which is consistent with previous work \citep[e.g.][]{Kordopatis15}. While the old, metal-poor thick disc stars have larger eccentricities and broader distribution, the mean eccentricities vary from 0.2 to larger than 0.5. For the thin disc stars, the eccentricity steady increases with stellar age and exhibits little change with [Fe/H] at a fixed age, possibly suggesting that the steady disc heating processes has played a decisive role to determine the orbital evolution of these stars \citep[e.g.][]{Ting19}. At a given age, the [Fe/H] of a star is an indicator of its Galactocentric radius because of the existence of radial [Fe/H] gradient for mono-age populations \citep[e.g.][]{Boeche13,Xiang15b,Xiang17a,Anders17,Wang19}. The lack of variations with [Fe/H] suggests that the overall eccentricity for stars with the same age has little radial variations in the range of our sample ($7<R<13$\,kpc). The spatial patterns of eccentricity in the $R$ -- $Z$ plane reflect mainly the superposition of stellar populations of different ages. The young, outer disc stars exhibit positive skewness. The patterns in both the age -- [Fe/H] and $R$ -- $Z$ plane are similar to the inverse cases for $V_{\phi}$ (see Figs.\,4 and 8). This is expected since the eccentricity is largely determined by $V_{\phi}$.
Interestingly, in 11 $<$ $R$ $<$ 13\,kpc, $-$0.25 $\lesssim$ $Z$ $\lesssim$ 0.25\,kpc, the dispersion and skewness of eccentricity exhibit peculiar behaviours, similar to the case of $V_{\phi}$. This is might be an effect caused by spiral arms (Wu et al. in preparation).

\begin{figure*}
\centering
\includegraphics[width=\linewidth]{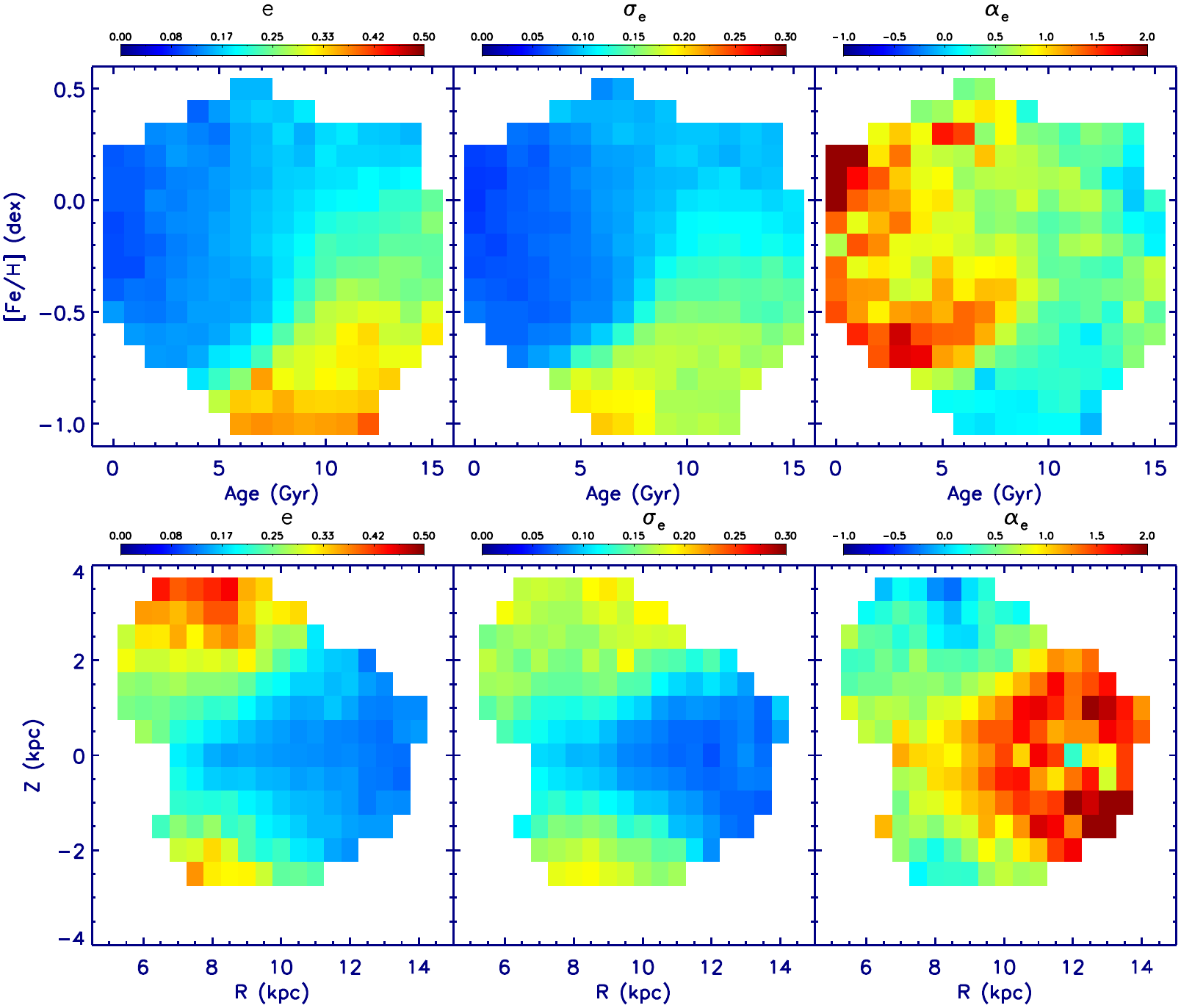}
\caption{Top panels: Colour-coded distributions of median ({\em left}), dispersion ({\em middle}), and skewness ({\em right}) of the eccentricity distribution functions for stars in individual age--[Fe/H] bins of 1\,Gyr $\times$ 0.1\,dex. Bottom panels: Colour-coded distributions of median (left), dispersion (middle), and skewness (right) of the eccentricity distribution functions for stars in individual $R$--$Z$ bins of 0.5\,kpc $\times$ 0.5\,kpc. \label{fig:fig1}}
\end{figure*}

Fig.\,11 plots the distributions of $Z_{max}$, $R_{g}$, $R_{peri}$, and $R_{apo}$. The figure clearly shows that the younger, more metal-rich stars have maximal orbital heights smaller than 1\,kpc, while the older, metal-poor stars have maximal orbital heights larger than 1\,kpc, giving credence to our definition of `thin' and `thick' disc populations in Section\,3. The thick disc stars have orbital guiding-centre radii of 4 -- 6\,kpc, peri-centre radii of 3 -- 6\,kpc, and apo-centre radii of $<10$\,kpc. This is consistent with the convention that thick disc stars are all formed in the inner disc \citep[e.g.][]{Haywood13}. It is clear that the thick disc stars exhibit smaller guiding-centre radii than those thin disc stars.
The figure also illustrates that, except for the young, metal-poor ones, stars in all age and [Fe/H] bins have apo-centre radii $\simeq$10\,kpc. In other words, the `inner' ($R_{\rm apo}<10$\,kpc) disc has complex stellar populations, which dominate most of age and [Fe/H] space, while the `outer' ($R_{\rm apo}>10$\,kpc) disc has a relatively simple formation history, as the stars dominate only a small, specific part of the age - [Fe/H] space.

\begin{figure*}
\centering
\includegraphics[width=\linewidth]{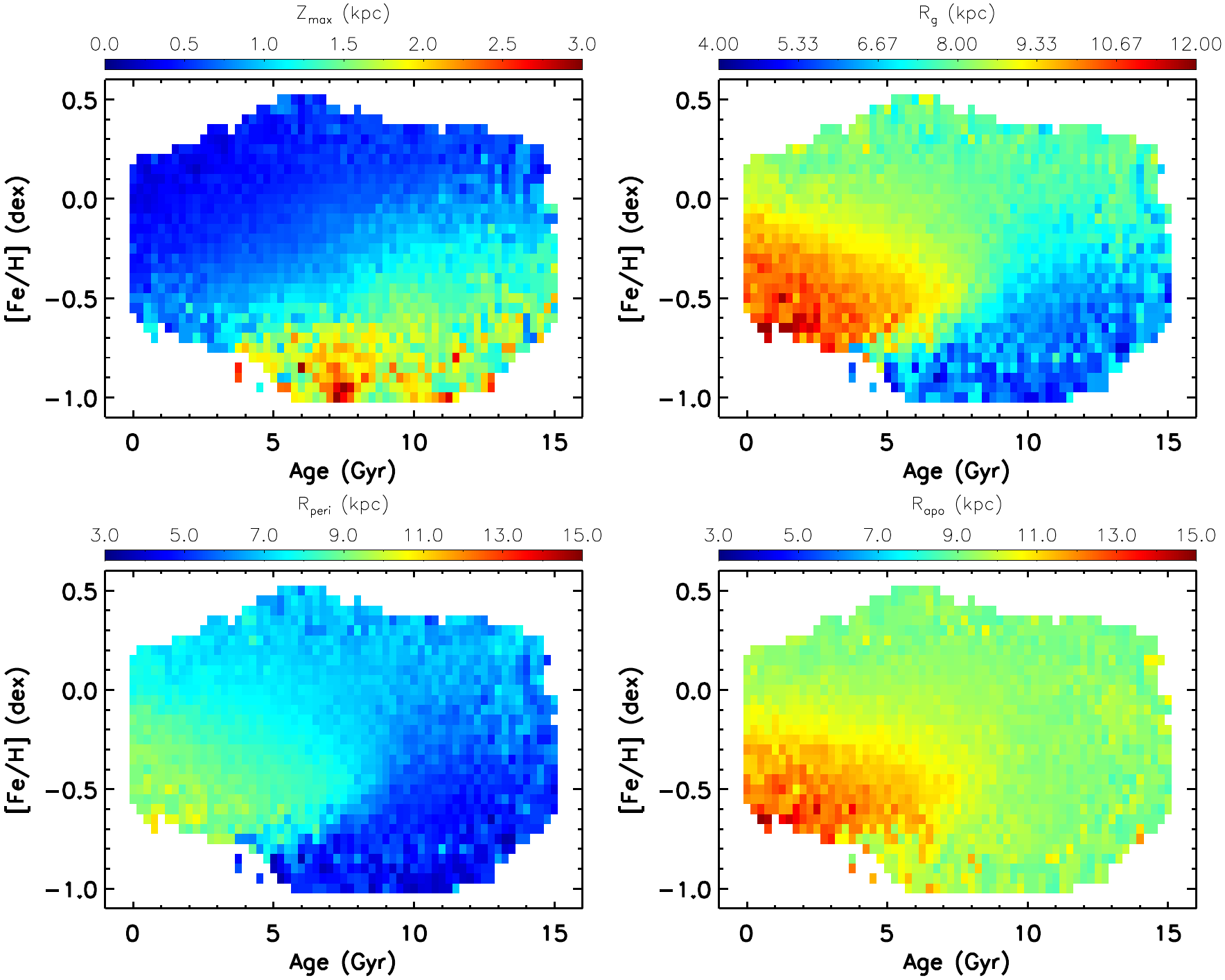}
\caption{Distributions of the maximal height $Z_{\rm max}$, guiding center radius $R_{g}$, pericenter radius $R_{\rm peri}$, and apocenter radius $R_{\rm apo}$ of the stellar orbits across the age--[Fe/H] plane. Colours represent the median values of the orbital parameters for stars in individual age--[Fe/H] bins of 0.25\,Gyr $\times$ 0.05\,dex. \label{fig:fig1}}
\end{figure*}

\subsection{Non-circular motion}

Due to non-circular nature of the stellar orbits, stars change their radial positions along the orbits without changing the angular momentum.
Here we study the impact of non-circular effect on the stellar distribution by comparing the current radial position ($R$) of the star with their orbital guiding-centre radii ($R_{g}$). Fig.\,12 shows the distributions of median value of $R-R_{g}$ for stars across the age -- metallicity and $R$ -- $Z$ planes. The figure shows a sharp border between the younger, metal-rich thin disc stars and the older, metal-poor thick disc stars: the thick disc stars show positive values of 1 -- 4\,kpc, indicating that they are currently at their apo-centre side. This is a strong non-circular effect considering the fact that 4\,kpc is comparable to their guiding-centre radii (Fig.\,11). However, the thin disc stars exhibit only small such effect, since the $R-R_{g}$ has a value of only $-$1\,kpc to $\sim$ 1\,kpc, depending on the location in the age--[Fe/H] and $R$ -- $Z$ planes. The young, metal-poor stars at the outermost disc ($R>12$\,kpc, $Z<0$\,kpc) exhibit a strongest negative $R-R_{g}$ of about $-1$\,kpc. Note that these stars belong to the warp structure of the outer disc in the third quadrant. The negative $R-R_{g}$ values suggest that these stars are currently at the peri-centre side of their orbits.

\begin{figure*}
\centering
\includegraphics[width=\linewidth]{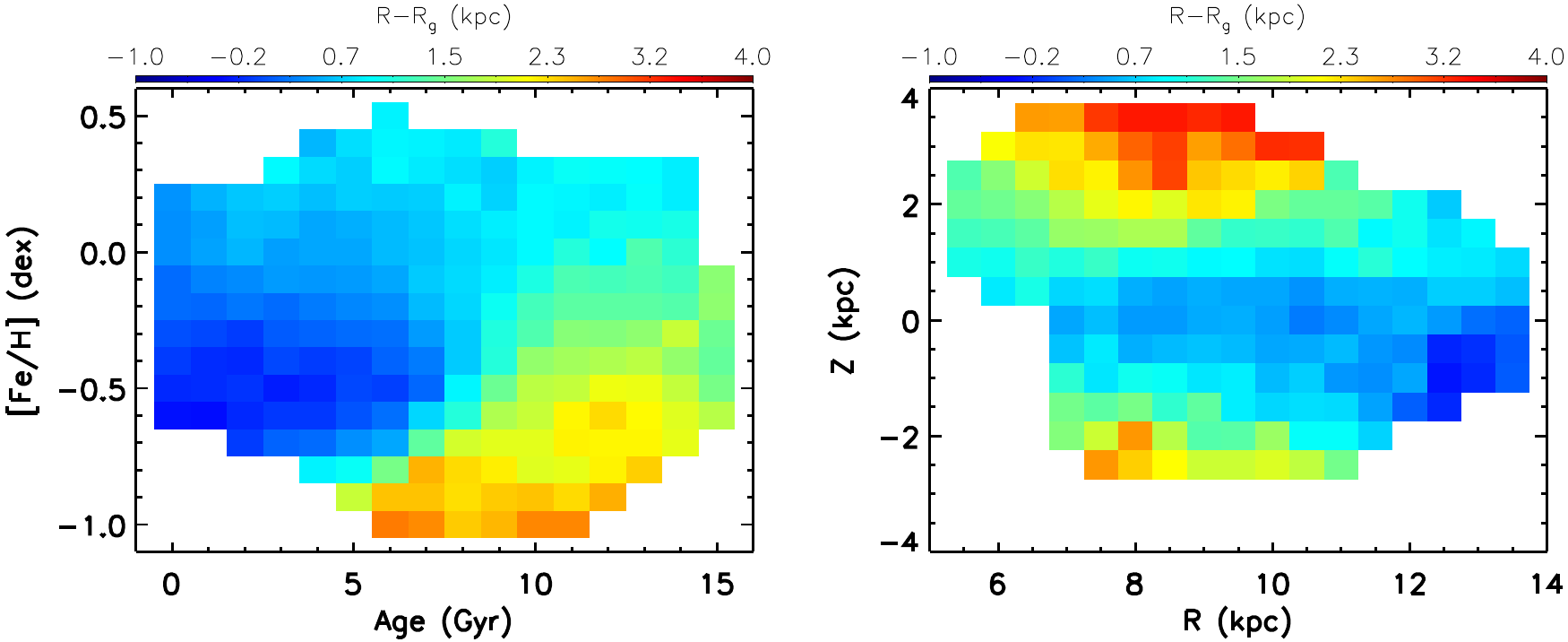}
\caption{Distributions of the $R - R_{\rm g}$ across the age--[Fe/H] ({\em left}) and the Galactic $R$--$Z$ ({\em right}) planes. Colours represent the median values of $R - R_{\rm g}$ for stars in individual age--[Fe/H] and $R$--$Z$ bins. \label{fig:fig1}}
\end{figure*}

\subsection{Churning}

Determining the stellar churning effect relies on knowledge about the stellar birth radii. This is a tough task, as the stellar kinematics themselves tell little about the birth position of a migrator. It has been suggested that, for some particular stellar populations, the stellar metallicity, as a fossil record for metallicity of the interstellar medium from which the star was born from, can be an indicator of the stellar birth place. For example, old ($\gtrsim8$\,Gyr), metal-rich (${\rm [Fe/H]}>0$) stars are generally expected to be formed at the inner part of the thin disc. So that if such stars occur at the solar neighbourhood or the outer disc with near circular orbits, they are likely arrived their current position via churning \citep[e.g.][]{Nordstrom04, Schonrich09, Kordopatis13, Boeche13, Chen19, Wang19}.

With the justifiable assumption that a negative radial metallicity gradient in the interstellar medium (ISM) is existed for most of the disc lifetime, and that the dispersion of the ISM metallicity at a given age and radius is small, \citet{Minchev18} proposed a method to determine the stellar birth radii based on their age and metallicity. Briefly, the method assumes that the ISM has a radial metallicity gradients of $-0.15$\,dex/kpc at the beginning of the disc formation, and monotonically flats to $-0.07$\,dex/kpc at present. The ISM metallicity at solar radius is assumed to vary from ${\rm [Fe/H]}\sim-1.0$\,dex at 13\,Gyr ago to 0.1\,dex at present. With these prior knowledge, the birth radius of a star then can be derived from their age and [Fe/H].

Furthermore, the effectiveness of this method has been verified by \citet{Chen19} via applying to the LAMOST stars. Similar to \citet{Chen19}, here we adopt the \citet{Minchev18} method to estimate the birth radii of our RGB stars. Note that the thick disc stars are widely suggested to exhibit an insignificant radial metallicity gradient, which is explained as an evidence that the disc stars are born from radially well-mixed gas \citep[e.g.][]{Cheng12, Boeche13,Haywood13, Xiang15a, Wang19}. The birth radii inferred from their age and metallicity might be problematic. Therefore we focus only on the thin disc stars, for which the method applies.

We select the thin disc stars through the [Fe/H]--[$\alpha$/Fe] plane. The distribution of our RGB stars in the [Fe/H] -- [$\alpha$/Fe] plane is shown in the left panel of Fig.\,13. The figure clearly displays two prominent sequences of stars that correspond to the high-$\alpha$ thick disc sequence and low-$\alpha$ thin disc sequence, respectively. We adopt the broken line shown in the Fig.\,13 as a criterion to distinguish the thin or thick disc stars. The right panel of Fig.\,13 suggests that the selected thin disc stars are located in the regions of the age -- [Fe/H] plane where the stars have small $J_R$ values, verifying the effectiveness of the selection criteria.

In Fig.\,14, we plot the distribution of the difference between the guiding-centre radii and the birth radii $R_{g}-R_{b}$ of the thin disc stars in age -- metallicity and $R$ -- $Z$ planes. The $R_{g}-R_{b}$ is a direct measurement of the effection of stellar redistribution due to radial migration. The figure suggests that the relatively old stars with super-solar metallicities show positive $R_{g}-R_{b}$ values, which means that they are outward migrators from the inner disc. While the younger, metal-poor stars show negative $R_{g}-R_{b}$ values, indicating that they are inward migrators from the outer disc. At a given age from 4 to 12\,Gyr, the maximal difference between $R_{g}$ and $R_{b}$ could reach 5\,kpc for the most metal-rich stars, indicating that these stars may have a minimal average migration speed of about 0.5 to 1\,kpc/Gyr, if we assume that the stars start to migrate since they are born. This is consistent with \citet{Quillen18} and \citet{Chen19}, who suggest that the stellar migration due to the excitation of the Galactic bar have a speed of 0.5--1\,kpc/Gyr, depending on the pitch angle of the bar. Interestingly, the data exhibits a lack of stars with ${\rm [Fe/H]}\gtrsim0.3$\,dex and age younger than 4\,Gyr. 
For the most metal-poor stars at age younger than 2\,Gyr, the distance of inward migration could reach 4--5\,kpc, suggesting a fast migration speed of larger than 2\,kpc/Gyr. We do not fully understand such a large migration speed. It is possible that such a fast migration speed might be related to some violent processes, such as perturbations of galaxy mergers. On the other hand, it is also possible that the $R_{b}$ estimates for these outer disc stars suffer large uncertainties. 

Fig.\,14 further shows the median distance alteration due to the radial migration for stars at different positions across the disc $R$--$Z$ plane. It suggests that, at around the solar radius ($7<R<9$\,kpc), the median distance alteration is only 0 -- 1\,kpc. For stars older than 8\,Gyr, the redistribution effect due to radial migration is much stronger, with a median value of 3\,kpc.
In the outer disc of 9$<$$R$$<$12\,kpc, the median distance alteration due to the stellar migration for the overall sample stars is small, from $-1$ to 0\,kpc, but the young stars shows a larger overall inward migration of $-$3\,kpc.

However, a small alteration in the radial position for the overall stars does not mean a weak migration effect. In fact, Fig.\,15 shows that, the stars with common birth radius show strong migrations. Particularly, for stars born at 12\,kpc, very few of them remain at their birth place, but most of them have migrated inward by a large distance ($>2$\,kpc). Such a net inward migration effect might be driven by splashes triggered by merger events of satellite galaxies at the outer disc. Particularly, the Figure shows an age stratification for these inward migrators, possibly an evidence suggesting that the migration in the outer disc was triggered by merger events that has been lasted in the past few giga years. This provides constraints on understanding the merger history of our Galaxy \citep[e.g.][]{Minchev16}.

The birth radius determination depends on assumptions on the time-dependent radial metallicity gradient of the ISM, in order to understand possible uncertainties raised from this assumption, we test a different ISM gradients by adopting the mono-age stellar metallicity gradients of \citet{Xiang15b}( see their Fig. 15) as an approximation of the ISM gradients. The comparison of the ISM metallicity adopted in this work for the birth radius calculation (i.e., from Minchev et al. 2018) and that of the \citet{Xiang15b} is shown in Fig.\,16. The figure also shows the comparison of the birth radius derived from these metallicity gradients, as well as the current position of $R_{b}$ $=$ 12\,kpc stars according to the \citet{Xiang15b} metallicity gradients. The results suggest that our conclusion about the migration is robust, i.e., the results are not significantly impacted by the adopted metallicity gradients. This is consistent with the findings of \citet{Feltzing20}, who showed that the birth radii of stars with different ages depends only marginally on the adopted ISM metallicity gradients, confirming the effectiveness of the \citet{Minchev18} method for estimating stellar birth radii.

\begin{figure*}
\centering
\includegraphics[width=\linewidth]{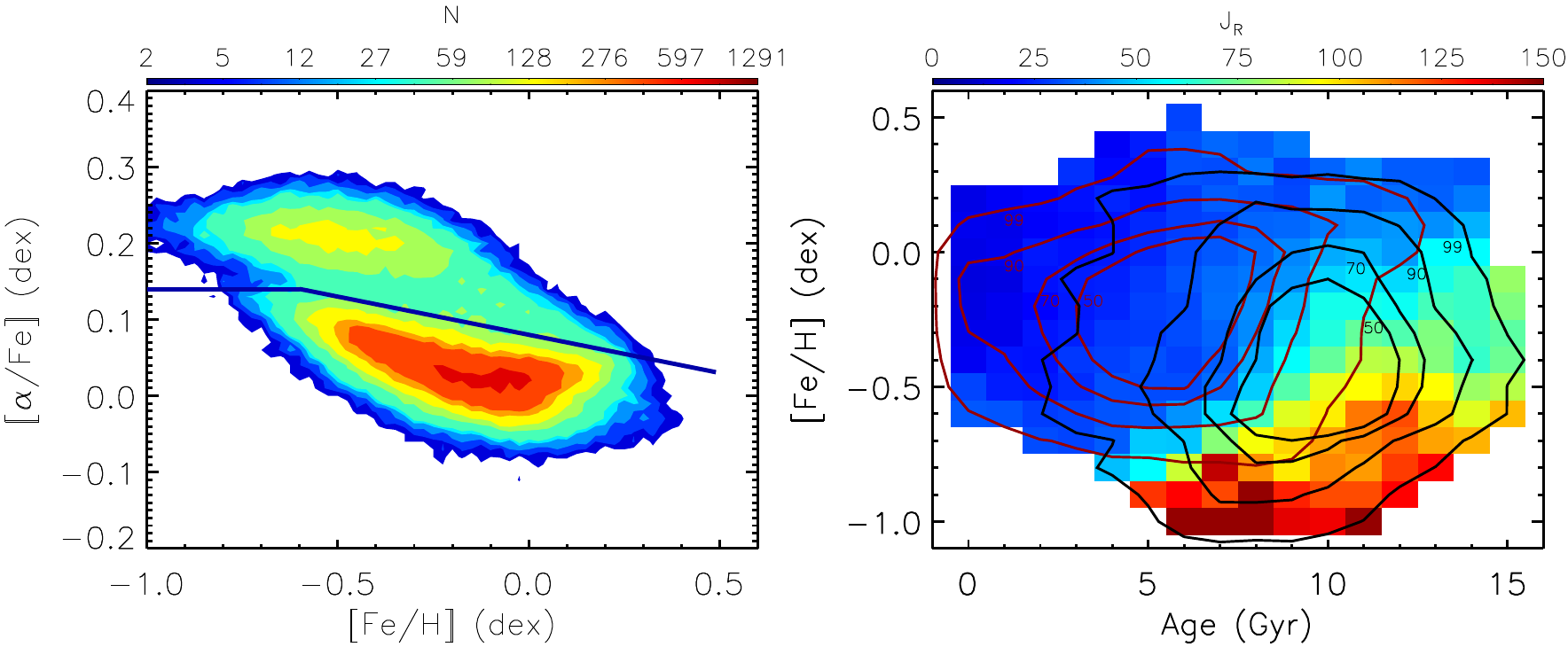}
\caption{{\it Left:} [$\alpha$/Fe] versus [Fe/H] diagram for our RGB sample. The blue solid line delineates the demarcation of the thin and thick disc sequences adopted by this work. {\it Right:} [Fe/H] versus Age diagram for our RGB sample stars, colour-coded by values of $J_{R}$. Also shown are the contours indicating 50,70,90, and 99 per cent of the selected thin (red) and thick disc (black) stars.\label{fig:fig1}}
\end{figure*}

\begin{figure*}
\centering
\includegraphics[width=\linewidth]{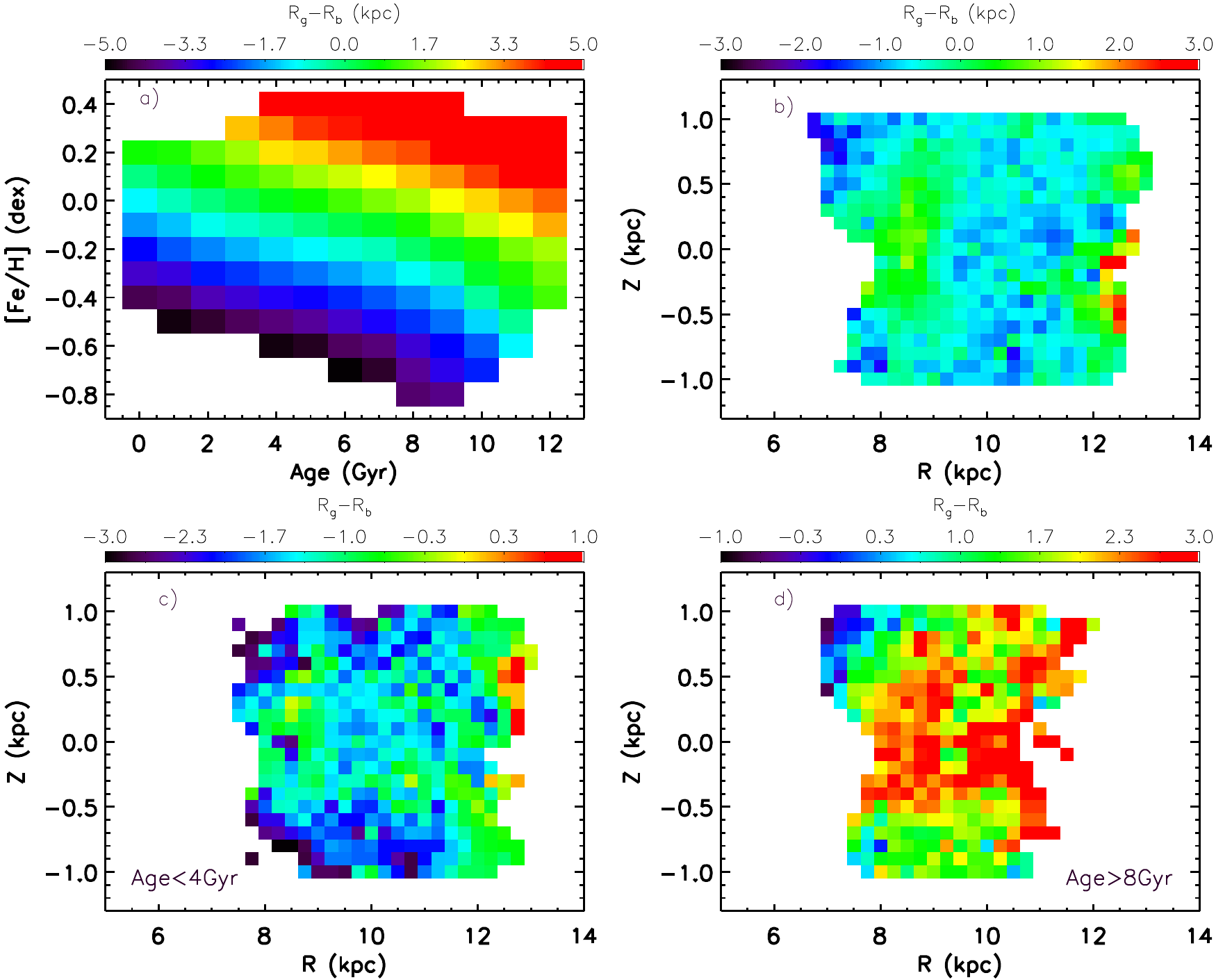}
\caption{Panel a), Colour-coded distribution of $R_{\rm g} - R_{\rm b}$ for the thin disc stars in the age--[Fe/H] plane. Panel b): Colour-coded distribution of $R_{\rm g} - R_{\rm b}$ for the thin disc stars in the Galactic $R$--$Z$ plane. Panel c): Same as Panel b), but for stars younger than 4\,Gyr. Panel d): Same as Panel c), but for stars older than 8. \label{fig:fig1}}
\end{figure*}

\begin{figure}
\centering
\includegraphics[width=90mm, angle=0]{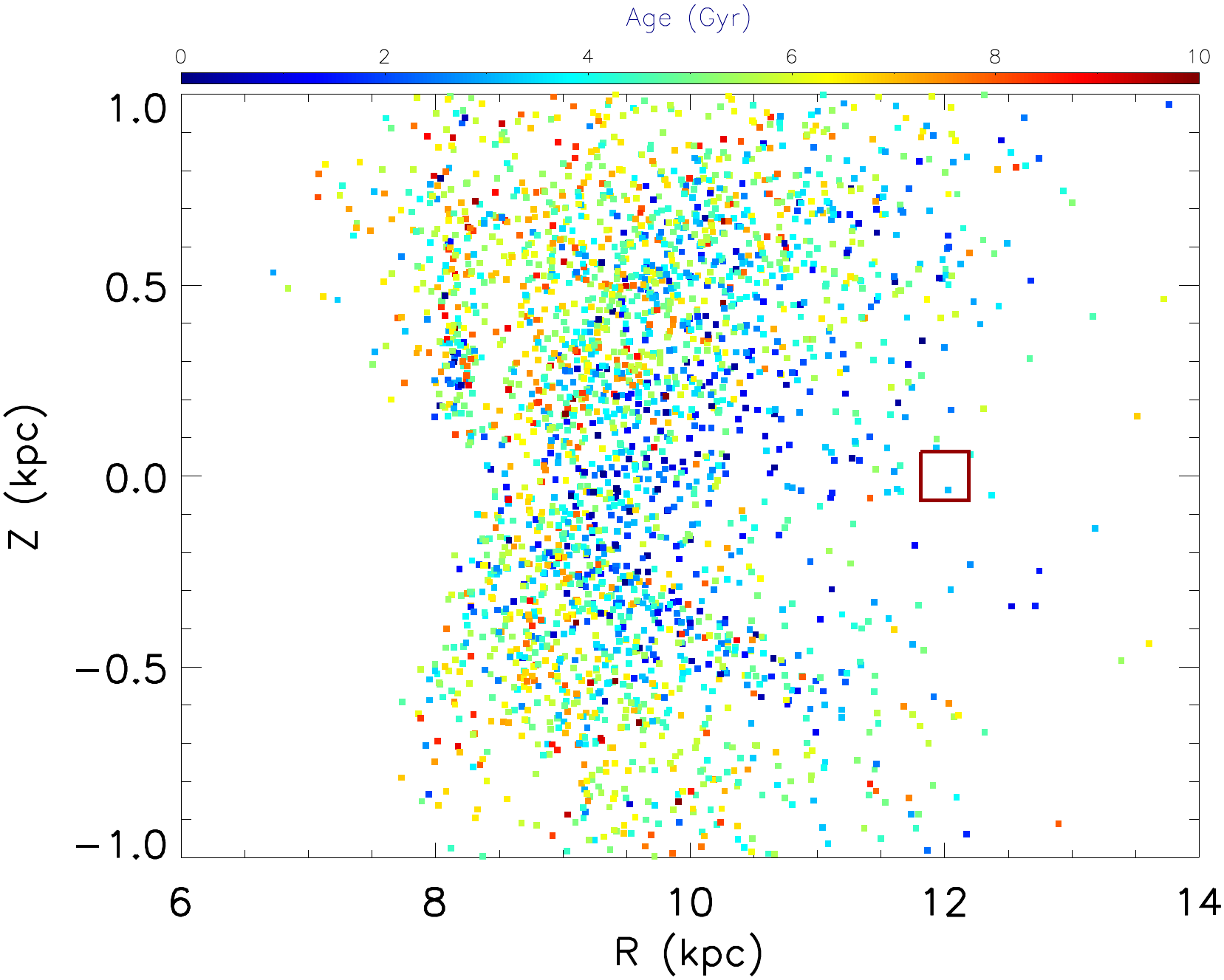}
\caption{Distributions in the $R$ -- $Z$ plane for stars born at 12\,kpc ($R_{\rm b}=12$\,kpc), as indicated by the red box. The color represents stellar age. Stratification effect in stellar age is visible. \label{fig:fig1}}
\end{figure}

\begin{figure*}
\centering
\includegraphics[width=\linewidth]{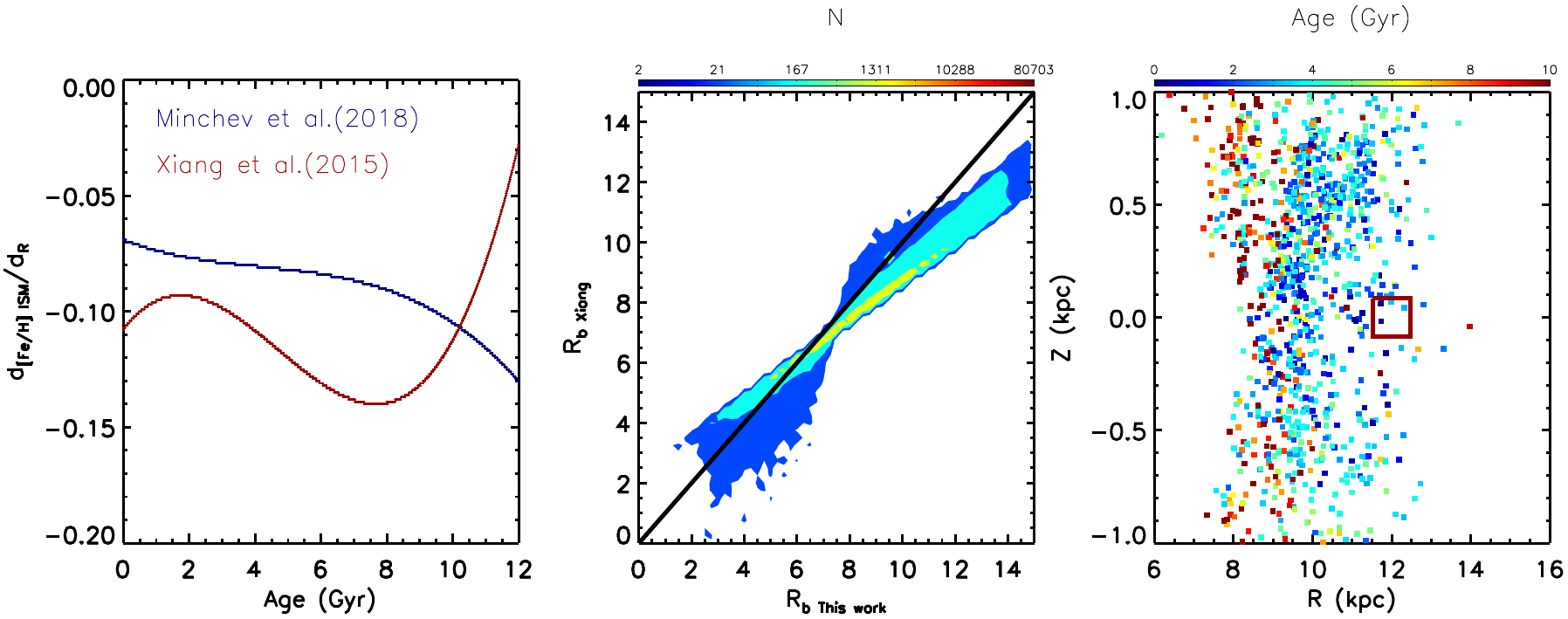}
\caption{{\it Left panel:} Comparison of the fits to the evolution functions of the metallicity gradient of \citet{Minchev18} with \citet{Xiang15b}(see their Fig.\,15). {\it In the middle panel:} Comparison of birth radius estimates of this work with metallicity gradient using of \citet{Xiang15b} (see their Fig.\,15). The black solid line represents the 1:1 line. The colour bar represent the number of stars. {\it Right panel:} Distributions in the $R$ -- $Z$ plane for stars born at 12\,kpc (birth radii are calculated with metallicity gradient using of \citet{Xiang15b}), as indicated by the red box. The colour bar presents stellar age. \label{fig:fig1}}
\end{figure*}

\section{Summary}

We have investigated the stellar kinematics of the Galactic disc in 7 $<$ $R$ $<$ 13\,kpc with a large sample of RGB stars combining LAMOST and Gaia. We characterize the median, dispersion and skewness of the distributions of the 3D stellar velocities, actions and orbital parameters across the age-metallicity and the disc $R$ -- $Z$ planes, and discussed the impact of stellar migration for the thin disc stars by comparing the guiding-centre
radii and the birth radii, which are derived based on their age and
[Fe/H]. The results reveal abundant kinematic trends and structures. Particularly, we find that:

i) The most prominent feature is the clear difference in the velocity and action distributions between the thin and thick disc stars. In the age--[Fe/H] plane, the kinematics exhibit a sharp border between the young, metal-rich thin disc stars and the old, metal-poor thick disc stars. This is particularly clear in the distributions of the actions, and presents in the median, dispersion, and skewness for all the 3D actions $J_{R}$, $J_{\phi}$, and $J_{Z}$. The kinematic distinction between the thin and thick stars is consistent with the chemical distinction in the [Fe/H]--[$\alpha$/Fe] plane. These results support the widely accepted conception that the Milky Way disc is composed of two major components which related to different formation and evolution history.

ii) The velocity distributions reveal several known kinematic substructures in the outer disc. Among of them, we confirm the existence of the north-south asymmetries for $V_{R}$, which present for stars of all ages and exhibit complicated patterns in the $R$ -- $Z$ planes. As a part of these complicated patterns, the high-velocity stream at $R\sim11$\,kpc and $Z\sim0.5$\,kpc is also shown in stars of all ages, and is particularly strong for young populations. The data also reveals the north-south asymmetries in $V_{Z}$ (also $V_{\phi}$) for $R>10$\,kpc, which is the kinematic feature of the disc warp in the outer disc.

iii) The skewness of the distributions of both $V_{\phi}$ and $J_{\phi}$ reveals a new substructure at $R\sim12$\,kpc and $Z\sim0$\,kpc, which is related to stars in between the spiral arms (the perseus arm and the outer arm) in the outer disc. The nature of this substructure deserves to be further studied.

iv) The thick disc stars have a typical guiding-centre radii smaller than 6\,kpc, indicating that they are mostly born in the inner disc of $R<6$\,kpc and reach their current positions via highly eccentric orbital motions. The thin disc stars, however, have much smaller eccentricities, which lead to typically only $\sim$1\,kpc difference between the guiding-centre radii and their current positions. These results are well consistent with former works \citep[e.g.][]{Haywood13,Kordopatis15,Jia18}.

v) The old ($>4$\,Gyr), metal-rich stars in our sample exhibit strong outward migration. At a given age, the most metal-rich stars are outward migrators born at 5\,kpc inner than their current positions, indicating a mean migration speed of about 0.5--1.0\,kpc/Gyr. While the metal-poor thin disc stars exhibit strong inward radial migration. At a given age, the most metal-poor stars might have been born at 3--5\,kpc further away from their current positions. For young stars, this suggests a fast mean migration speed, probably triggered by merge events.

vi) All stars born at the outer disc of $R\gtrsim12$\,kpc may have experienced strong inward migration. The spatial distribution in the $R$ -- $Z$ plane for the inward migrators born at a given distance (e.g., 12\,kpc) shows an age stratification, possibly an evidence that these inward migrators are consequences of splashes triggered by merger events of satellite galaxies that have lasted in the past few giga years.

\section{Data availability}

The data underlying this article will be shared on reasonable request to the corresponding author.

\section*{Acknowledgments}

It is a pleasure to thank the anonymous referee and editior for their helpful
suggestions. This work is supported by the National Natural Science Foundation of China under grant No.11988101, 11903044, 11890694, 11625313, the Joint Research Fund in Astronomy (U2031203), National Key R\&D Program of China No. 2019YFA0405502, M.-S. Xiang and Y. Huang acknowledge supports from NSFC Grant No. 11703035. Cultivation Project for LAMOST Scientific Payoff and Research Achievement of CAMS-CAS.
The LAMOST FELLOWSHIP is supported by Special Funding for Advanced Users, budgeted and administrated by Center for Astronomical Mega-Science, Chinese Academy of Sciences (CAMS).
Guoshoujing Telescope (the Large Sky Area Multi-Object Fiber Spectroscopic Telescope LAMOST)
is a National Major Scientific Project built by the Chinese Academy of Sciences.
Funding for the project has been provided by the National Development and Reform Commission.
LAMOST is operated and managed by the National Astronomical Observatories, Chinese Academy of Sciences.
This work has also made use of data from the European Space Agency (ESA) mission
Gaia \footnote{https://www.cosmos.esa.int/gaia}, processed
by the Gaia Data Processing and Analysis Consortium
(DPAC, \footnote{https://www.cosmos.esa.int/web/gaia/dpac/consortium}). Funding for the DPAC has been provided
by national institutions, in particular the institutions
participating in the Gaia Multilateral Agreement.

\bibliographystyle{mnras}
\bibliography{correlation_clean1bib}

\begin{thebibliography}{}
\makeatletter
\relax
\def\mn@urlcharsother{\let\do\@makeother \do\$\do\&\do\#\do\^\do\_\do\%\do\~}
\def\mn@doi{\begingroup\mn@urlcharsother \@ifnextchar [ {\mn@doi@}
  {\mn@doi@[]}}
\def\mn@doi@[#1]#2{\def\@tempa{#1}\ifx\@tempa\@empty \href
  {http://dx.doi.org/#2} {doi:#2}\else \href {http://dx.doi.org/#2} {#1}\fi
  \endgroup}
\def\mn@eprint#1#2{\mn@eprint@#1:#2::\@nil}
\def\mn@eprint@arXiv#1{\href {http://arxiv.org/abs/#1} {{\tt arXiv:#1}}}
\def\mn@eprint@dblp#1{\href {http://dblp.uni-trier.de/rec/bibtex/#1.xml}
  {dblp:#1}}
\def\mn@eprint@#1:#2:#3:#4\@nil{\def\@tempa {#1}\def\@tempb {#2}\def\@tempc
  {#3}\ifx \@tempc \@empty \let \@tempc \@tempb \let \@tempb \@tempa \fi \ifx
  \@tempb \@empty \def\@tempb {arXiv}\fi \@ifundefined
  {mn@eprint@\@tempb}{\@tempb:\@tempc}{\expandafter \expandafter \csname
  mn@eprint@\@tempb\endcsname \expandafter{\@tempc}}}

\bibitem[\protect\citeauthoryear{{Adibekyan} et~al.,}{{Adibekyan}
  et~al.}{2013}]{Adibekyan13}
{Adibekyan} V.~Z.,  et~al., 2013, \mn@doi [A\&A] {10.1051/0004-6361/201321520},
  \href {https://ui.adsabs.harvard.edu/abs/2013A&A...554A..44A} {554, A44}

\bibitem[\protect\citeauthoryear{{Allende Prieto}, {Kawata}  \&
  {Cropper}}{{Allende Prieto} et~al.}{2016}]{Allende16}
{Allende Prieto} C.,  {Kawata} D.,   {Cropper} M.,  2016, \mn@doi [\aap]
  {10.1051/0004-6361/201629787}, \href
  {https://ui.adsabs.harvard.edu/abs/2016A&A...596A..98A} {596, A98}

\bibitem[\protect\citeauthoryear{{Anders} et~al.,}{{Anders}
  et~al.}{2017}]{Anders17}
{Anders} F.,  et~al., 2017, \mn@doi [\aap] {10.1051/0004-6361/201629363}, \href
  {https://ui.adsabs.harvard.edu/abs/2017A&A...600A..70A} {600, A70}

\bibitem[\protect\citeauthoryear{{Anders}, {Chiappini}, {Santiago},
  {Matijevi{\v{c}}}, {Queiroz}, {Steinmetz}  \& {Guiglion}}{{Anders}
  et~al.}{2018}]{A18}
{Anders} F.,  {Chiappini} C.,  {Santiago} B.~X.,  {Matijevi{\v{c}}} G.,
  {Queiroz} A.~B.,  {Steinmetz} M.,   {Guiglion} G.,  2018, \mn@doi [\aap]
  {10.1051/0004-6361/201833099}, \href
  {https://ui.adsabs.harvard.edu/abs/2018A&A...619A.125A} {619, A125}

\bibitem[\protect\citeauthoryear{{Antoja} et~al.,}{{Antoja}
  et~al.}{2018}]{Antoja18}
{Antoja} T.,  et~al., 2018, \mn@doi [\nat] {10.1038/s41586-018-0510-7}, \href
  {https://ui.adsabs.harvard.edu/abs/2018Natur.561..360A} {561, 360}

\bibitem[\protect\citeauthoryear{{Aumer} \& {Binney}}{{Aumer} \&
  {Binney}}{2009}]{Aumer09}
{Aumer} M.,  {Binney} J.~J.,  2009, \mn@doi [\mnras]
  {10.1111/j.1365-2966.2009.15053.x}, \href
  {https://ui.adsabs.harvard.edu/abs/2009MNRAS.397.1286A} {397, 1286}

\bibitem[\protect\citeauthoryear{{Aumer} \& {Binney}}{{Aumer} \&
  {Binney}}{2017}]{Aumer17}
{Aumer} M.,  {Binney} J.,  2017, \mn@doi [\mnras] {10.1093/mnras/stx1300},
  \href {https://ui.adsabs.harvard.edu/abs/2017MNRAS.470.2113A} {470, 2113}

\bibitem[\protect\citeauthoryear{{Aumer}, {Binney}  \& {Sch{\"o}nrich}}{{Aumer}
  et~al.}{2017}]{Aumer17b}
{Aumer} M.,  {Binney} J.,   {Sch{\"o}nrich} R.,  2017, \mn@doi [\mnras]
  {10.1093/mnras/stx1483}, \href
  {https://ui.adsabs.harvard.edu/abs/2017MNRAS.470.3685A} {470, 3685}

\bibitem[\protect\citeauthoryear{{Bailer-Jones}, {Rybizki}, {Fouesneau},
  {Mantelet}  \& {Andrae}}{{Bailer-Jones} et~al.}{2018}]{Bailer18}
{Bailer-Jones} C.~A.~L.,  {Rybizki} J.,  {Fouesneau} M.,  {Mantelet} G.,
  {Andrae} R.,  2018, \mn@doi [\aj] {10.3847/1538-3881/aacb21}, \href
  {https://ui.adsabs.harvard.edu/abs/2018AJ....156...58B} {156, 58}

\bibitem[\protect\citeauthoryear{{Bensby}, {Feltzing}  \&
  {Lundstr{\"o}m}}{{Bensby} et~al.}{2003}]{Bensby03}
{Bensby} T.,  {Feltzing} S.,   {Lundstr{\"o}m} I.,  2003, \mn@doi [\aap]
  {10.1051/0004-6361:20031213}, \href
  {https://ui.adsabs.harvard.edu/abs/2003A&A...410..527B} {410, 527}

\bibitem[\protect\citeauthoryear{{Binney} \& {Tremaine}}{{Binney} \&
  {Tremaine}}{2008}]{Binney08}
{Binney} J.,  {Tremaine} S.,  2008, {Galactic Dynamics: Second Edition}

\bibitem[\protect\citeauthoryear{{Bland-Hawthorn} et~al.,}{{Bland-Hawthorn}
  et~al.}{2019}]{Bland19}
{Bland-Hawthorn} J.,  et~al., 2019, \mn@doi [\mnras] {10.1093/mnras/stz217},
  \href {https://ui.adsabs.harvard.edu/abs/2019MNRAS.486.1167B} {486, 1167}

\bibitem[\protect\citeauthoryear{{Boeche} et~al.,}{{Boeche}
  et~al.}{2013}]{Boeche13}
{Boeche} C.,  et~al., 2013, \mn@doi [\aap] {10.1051/0004-6361/201322085}, \href
  {https://ui.adsabs.harvard.edu/abs/2013A&A...559A..59B} {559, A59}

\bibitem[\protect\citeauthoryear{{Bovy}}{{Bovy}}{2015}]{Bovy15}
{Bovy} J.,  2015, \mn@doi [\apjs] {10.1088/0067-0049/216/2/29}, \href
  {https://ui.adsabs.harvard.edu/abs/2015ApJS..216...29B} {216, 29}

\bibitem[\protect\citeauthoryear{{Bovy}, {Rix}, {Liu}, {Hogg}, {Beers}  \&
  {Lee}}{{Bovy} et~al.}{2012a}]{Bovy12b}
{Bovy} J.,  {Rix} H.-W.,  {Liu} C.,  {Hogg} D.~W.,  {Beers} T.~C.,   {Lee}
  Y.~S.,  2012a, \mn@doi [\apj] {10.1088/0004-637X/753/2/148}, \href
  {https://ui.adsabs.harvard.edu/abs/2012ApJ...753..148B} {753, 148}

\bibitem[\protect\citeauthoryear{{Bovy}, {Rix}, {Hogg}, {Beers}, {Lee}  \&
  {Zhang}}{{Bovy} et~al.}{2012b}]{Bovy12}
{Bovy} J.,  {Rix} H.-W.,  {Hogg} D.~W.,  {Beers} T.~C.,  {Lee} Y.~S.,   {Zhang}
  L.,  2012b, \mn@doi [\apj] {10.1088/0004-637X/755/2/115}, \href
  {https://ui.adsabs.harvard.edu/abs/2012ApJ...755..115B} {755, 115}

\bibitem[\protect\citeauthoryear{{Carlin} et~al.,}{{Carlin}
  et~al.}{2013}]{Carlin13}
{Carlin} J.~L.,  et~al., 2013, \mn@doi [\apjl] {10.1088/2041-8205/777/1/L5},
  \href {https://ui.adsabs.harvard.edu/abs/2013ApJ...777L...5C} {777, L5}

\bibitem[\protect\citeauthoryear{{Carrillo} et~al.,}{{Carrillo}
  et~al.}{2018}]{Carrillo18}
{Carrillo} I.,  et~al., 2018, \mn@doi [\mnras] {10.1093/mnras/stx3342}, \href
  {https://ui.adsabs.harvard.edu/abs/2018MNRAS.475.2679C} {475, 2679}

\bibitem[\protect\citeauthoryear{{Chen}, {Wang}, {Deng}, {de Grijs}, {Liu}  \&
  {Tian}}{{Chen} et~al.}{2019a}]{ChenXD19}
{Chen} X.,  {Wang} S.,  {Deng} L.,  {de Grijs} R.,  {Liu} C.,   {Tian} H.,
  2019a, \mn@doi [Nature Astronomy] {10.1038/s41550-018-0686-7}, \href
  {https://ui.adsabs.harvard.edu/abs/2019NatAs...3..320C} {3, 320}

\bibitem[\protect\citeauthoryear{{Chen}, {Zhao}, {Zhao}, {Liang}, {Wu}, {Jia},
  {Tian}  \& {Liu}}{{Chen} et~al.}{2019b}]{Chen19}
{Chen} Y.~Q.,  {Zhao} G.,  {Zhao} J.~K.,  {Liang} X.~L.,  {Wu} Y.~Q.,  {Jia}
  Y.~P.,  {Tian} H.,   {Liu} J.~M.,  2019b, \mn@doi [\aj]
  {10.3847/1538-3881/ab5283}, \href
  {https://ui.adsabs.harvard.edu/abs/2019AJ....158..249C} {158, 249}

\bibitem[\protect\citeauthoryear{{Cheng} et~al.,}{{Cheng}
  et~al.}{2012}]{Cheng12}
{Cheng} J.~Y.,  et~al., 2012, \mn@doi [\apj] {10.1088/0004-637X/746/2/149},
  \href {https://ui.adsabs.harvard.edu/abs/2012ApJ...746..149C} {746, 149}

\bibitem[\protect\citeauthoryear{{Chequers}, {Widrow}  \& {Darling}}{{Chequers}
  et~al.}{2018}]{Chequers18}
{Chequers} M.~H.,  {Widrow} L.~M.,   {Darling} K.,  2018, \mn@doi [\mnras]
  {10.1093/mnras/sty2114}, \href
  {https://ui.adsabs.harvard.edu/abs/2018MNRAS.480.4244C} {480, 4244}

\bibitem[\protect\citeauthoryear{{Chiba} \& {Beers}}{{Chiba} \&
  {Beers}}{2000}]{Chiba00}
{Chiba} M.,  {Beers} T.~C.,  2000, \mn@doi [\aj] {10.1086/301409}, \href
  {https://ui.adsabs.harvard.edu/abs/2000AJ....119.2843C} {119, 2843}

\bibitem[\protect\citeauthoryear{{Coronado}, {Rix}, {Trick}, {El-Badry},
  {Rybizki}  \& {Xiang}}{{Coronado} et~al.}{2020}]{Coronado20}
{Coronado} J.,  {Rix} H.-W.,  {Trick} W.~H.,  {El-Badry} K.,  {Rybizki} J.,
  {Xiang} M.,  2020, \mn@doi [\mnras] {10.1093/mnras/staa1358}, \href
  {https://ui.adsabs.harvard.edu/abs/2020MNRAS.495.4098C} {495, 4098}

\bibitem[\protect\citeauthoryear{{Curir}, {Mazzei}  \& {Murante}}{{Curir}
  et~al.}{2004}]{Curir04}
{Curir} A.,  {Mazzei} P.,   {Murante} G.,  2004, {Disks evolution in a
  Cosmological framework}.
pp 335--339, \mn@doi{10.1007/978-1-4020-2862-5_31}

\bibitem[\protect\citeauthoryear{{Curir}, {Lattanzi}, {Spagna}, {Matteucci},
  {Murante}, {Re Fiorentin}  \& {Spitoni}}{{Curir} et~al.}{2012}]{Curir12}
{Curir} A.,  {Lattanzi} M.~G.,  {Spagna} A.,  {Matteucci} F.,  {Murante} G.,
  {Re Fiorentin} P.,   {Spitoni} E.,  2012, \mn@doi [\aap]
  {10.1051/0004-6361/201118558}, \href
  {https://ui.adsabs.harvard.edu/abs/2012A&A...545A.133C} {545, A133}

\bibitem[\protect\citeauthoryear{{De Silva} et~al.,}{{De Silva}
  et~al.}{2015}]{De15}
{De Silva} G.~M.,  et~al., 2015, \mn@doi [\mnras] {10.1093/mnras/stv327}, \href
  {https://ui.adsabs.harvard.edu/abs/2015MNRAS.449.2604D} {449, 2604}

\bibitem[\protect\citeauthoryear{{Deng} et~al.,}{{Deng} et~al.}{2012}]{Deng12}
{Deng} L.-C.,  et~al., 2012, \mn@doi [Research in Astronomy and Astrophysics]
  {10.1088/1674-4527/12/7/003}, \href
  {https://ui.adsabs.harvard.edu/abs/2012RAA....12..735D} {12, 735}

\bibitem[\protect\citeauthoryear{{Dobbie} \& {Warren}}{{Dobbie} \&
  {Warren}}{2020}]{Dobbie20}
{Dobbie} P.~S.,  {Warren} S.~J.,  2020, \mn@doi [The Open Journal of
  Astrophysics] {10.21105/astro.2003.05757}, \href
  {https://ui.adsabs.harvard.edu/abs/2020OJAp....3E...5D} {3, 5}

\bibitem[\protect\citeauthoryear{{Eilers}, {Hogg}, {Rix}, {Frankel}, {Hunt},
  {Fouvry}  \& {Buck}}{{Eilers} et~al.}{2020}]{Eilers20}
{Eilers} A.-C.,  {Hogg} D.~W.,  {Rix} H.-W.,  {Frankel} N.,  {Hunt} J. A.~S.,
  {Fouvry} J.-B.,   {Buck} T.,  2020, arXiv e-prints, \href
  {https://ui.adsabs.harvard.edu/abs/2020arXiv200301132E} {p. arXiv:2003.01132}

\bibitem[\protect\citeauthoryear{{Feltzing}, {Bowers}  \& {Agertz}}{{Feltzing}
  et~al.}{2020}]{Feltzing20}
{Feltzing} S.,  {Bowers} J.~B.,   {Agertz} O.,  2020, \mn@doi [\mnras]
  {10.1093/mnras/staa340}, \href
  {https://ui.adsabs.harvard.edu/abs/2020MNRAS.493.1419F} {493, 1419}

\bibitem[\protect\citeauthoryear{{Feuillet}, {Frankel}, {Lind}, {Frinchaboy},
  {Garc{\'\i}a-Hern{\'a}ndez}, {Lane}, {Nitschelm}  \&
  {Roman-Lopes}}{{Feuillet} et~al.}{2019}]{Feuillet19}
{Feuillet} D.~K.,  {Frankel} N.,  {Lind} K.,  {Frinchaboy} P.~M.,
  {Garc{\'\i}a-Hern{\'a}ndez} D.~A.,  {Lane} R.~R.,  {Nitschelm} C.,
  {Roman-Lopes} A.~r.,  2019, \mn@doi [\mnras] {10.1093/mnras/stz2221}, \href
  {https://ui.adsabs.harvard.edu/abs/2019MNRAS.489.1742F} {489, 1742}

\bibitem[\protect\citeauthoryear{{Frankel}, {Rix}, {Ting}, {Ness}  \&
  {Hogg}}{{Frankel} et~al.}{2018}]{Frankel18}
{Frankel} N.,  {Rix} H.-W.,  {Ting} Y.-S.,  {Ness} M.,   {Hogg} D.~W.,  2018,
  \mn@doi [\apj] {10.3847/1538-4357/aadba5}, \href
  {https://ui.adsabs.harvard.edu/abs/2018ApJ...865...96F} {865, 96}

\bibitem[\protect\citeauthoryear{{Freeman} \& {Bland-Hawthorn}}{{Freeman} \&
  {Bland-Hawthorn}}{2002}]{Freeman02}
{Freeman} K.,  {Bland-Hawthorn} J.,  2002, \mn@doi [\araa]
  {10.1146/annurev.astro.40.060401.093840}, \href
  {https://ui.adsabs.harvard.edu/abs/2002ARA&A..40..487F} {40, 487}

\bibitem[\protect\citeauthoryear{{Gaia Collaboration} et~al.,}{{Gaia
  Collaboration} et~al.}{2016a}]{Gaia16}
{Gaia Collaboration} et~al., 2016a, \mn@doi [\aap]
  {10.1051/0004-6361/201629272}, \href
  {https://ui.adsabs.harvard.edu/abs/2016A&A...595A...1G} {595, A1}

\bibitem[\protect\citeauthoryear{{Gaia Collaboration} et~al.,}{{Gaia
  Collaboration} et~al.}{2016b}]{Brown16}
{Gaia Collaboration} et~al., 2016b, \mn@doi [\aap]
  {10.1051/0004-6361/201629512}, \href
  {https://ui.adsabs.harvard.edu/abs/2016A&A...595A...2G} {595, A2}

\bibitem[\protect\citeauthoryear{{Gaia Collaboration} et~al.,}{{Gaia
  Collaboration} et~al.}{2018}]{Gaia18}
{Gaia Collaboration} et~al., 2018, \mn@doi [\aap]
  {10.1051/0004-6361/201833051}, \href
  {https://ui.adsabs.harvard.edu/abs/2018A&A...616A...1G} {616, A1}

\bibitem[\protect\citeauthoryear{{Gilmore} \& {Reid}}{{Gilmore} \&
  {Reid}}{1983}]{Gilmore83}
{Gilmore} G.,  {Reid} N.,  1983, \mn@doi [\mnras] {10.1093/mnras/202.4.1025},
  \href {https://ui.adsabs.harvard.edu/abs/1983MNRAS.202.1025G} {202, 1025}

\bibitem[\protect\citeauthoryear{{Grand} \& {Kawata}}{{Grand} \&
  {Kawata}}{2016}]{Grand16}
{Grand} R.~J.~J.,  {Kawata} D.,  2016, \mn@doi [Astronomische Nachrichten]
  {10.1002/asna.201612407}, \href
  {https://ui.adsabs.harvard.edu/abs/2016AN....337..957G} {337, 957}

\bibitem[\protect\citeauthoryear{{Grieves} et~al.,}{{Grieves}
  et~al.}{2018}]{Grieves18}
{Grieves} N.,  et~al., 2018, \mn@doi [\mnras] {10.1093/mnras/sty2431}, \href
  {https://ui.adsabs.harvard.edu/abs/2018MNRAS.481.3244G} {481, 3244}

\bibitem[\protect\citeauthoryear{{Hayden} et~al.,}{{Hayden}
  et~al.}{2014}]{Hayden14}
{Hayden} M.~R.,  et~al., 2014, \mn@doi [\aj] {10.1088/0004-6256/147/5/116},
  \href {https://ui.adsabs.harvard.edu/abs/2014AJ....147..116H} {147, 116}

\bibitem[\protect\citeauthoryear{{Hayden}, {Recio-Blanco}, {de Laverny},
  {Mikolaitis}  \& {Worley}}{{Hayden} et~al.}{2017}]{Hayden17}
{Hayden} M.~R.,  {Recio-Blanco} A.,  {de Laverny} P.,  {Mikolaitis} S.,
  {Worley} C.~C.,  2017, \mn@doi [\aap] {10.1051/0004-6361/201731494}, \href
  {https://ui.adsabs.harvard.edu/abs/2017A&A...608L...1H} {608, L1}

\bibitem[\protect\citeauthoryear{{Hayden} et~al.,}{{Hayden}
  et~al.}{2020}]{Hayden20}
{Hayden} M.~R.,  et~al., 2020, \mn@doi [\mnras] {10.1093/mnras/staa335}, \href
  {https://ui.adsabs.harvard.edu/abs/2020MNRAS.493.2952H} {493, 2952}

\bibitem[\protect\citeauthoryear{{Haywood}, {Di Matteo}, {Lehnert}, {Katz}  \&
  {G{\'o}mez}}{{Haywood} et~al.}{2013}]{Haywood13}
{Haywood} M.,  {Di Matteo} P.,  {Lehnert} M.~D.,  {Katz} D.,   {G{\'o}mez} A.,
  2013, \mn@doi [\aap] {10.1051/0004-6361/201321397}, \href
  {https://ui.adsabs.harvard.edu/abs/2013A&A...560A.109H} {560, A109}

\bibitem[\protect\citeauthoryear{{Holmberg}, {Nordstr{\"o}m}  \&
  {Andersen}}{{Holmberg} et~al.}{2009}]{Holmberg09}
{Holmberg} J.,  {Nordstr{\"o}m} B.,   {Andersen} J.,  2009, \mn@doi [\aap]
  {10.1051/0004-6361/200811191}, \href
  {https://ui.adsabs.harvard.edu/abs/2009A&A...501..941H} {501, 941}

\bibitem[\protect\citeauthoryear{{Huang}, {Liu}, {Yuan}, {Xiang}, {Chen}  \&
  {Zhang}}{{Huang} et~al.}{2015}]{Huang15}
{Huang} Y.,  {Liu} X.-W.,  {Yuan} H.-B.,  {Xiang} M.-S.,  {Chen} B.-Q.,
  {Zhang} H.-W.,  2015, \mn@doi [\mnras] {10.1093/mnras/stv1991}, \href
  {http://adsabs.harvard.edu/abs/2015MNRAS.454.2863H} {454, 2863}

\bibitem[\protect\citeauthoryear{{Huang} et~al.,}{{Huang}
  et~al.}{2018}]{Huang18}
{Huang} Y.,  et~al., 2018, \mn@doi [\apj] {10.3847/1538-4357/aad285}, \href
  {https://ui.adsabs.harvard.edu/abs/2018ApJ...864..129H} {864, 129}

\bibitem[\protect\citeauthoryear{{Jia}, {Chen}, {Zhao}, {Xue}, {Zhao}, {Yang}
  \& {Li}}{{Jia} et~al.}{2018}]{Jia18}
{Jia} Y.,  {Chen} Y.,  {Zhao} G.,  {Xue} X.,  {Zhao} J.,  {Yang} C.,   {Li} C.,
   2018, \mn@doi [\apj] {10.3847/1538-4357/aad3bb}, \href
  {https://ui.adsabs.harvard.edu/abs/2018ApJ...863...93J} {863, 93}

\bibitem[\protect\citeauthoryear{{Jing} et~al.,}{{Jing} et~al.}{2016}]{Jing16}
{Jing} Y.,  et~al., 2016, \mn@doi [\mnras] {10.1093/mnras/stw2230}, \href
  {https://ui.adsabs.harvard.edu/abs/2016MNRAS.463.3390J} {463, 3390}

\bibitem[\protect\citeauthoryear{{Katz} et~al.,}{{Katz} et~al.}{2018}]{Katz18}
{Katz} D.,  et~al., 2018, \mn@doi [\aap] {10.1051/0004-6361/201832865}, \href
  {https://ui.adsabs.harvard.edu/abs/2018A&A...616A..11G} {616, A11}

\bibitem[\protect\citeauthoryear{{Katz} et~al.,}{{Katz} et~al.}{2019}]{Katz19}
{Katz} D.,  et~al., 2019, \mn@doi [\aap] {10.1051/0004-6361/201833273}, \href
  {https://ui.adsabs.harvard.edu/abs/2019A&A...622A.205K} {622, A205}

\bibitem[\protect\citeauthoryear{{Kawata}, {Grand}, {Gibson}, {Casagrande},
  {Hunt}  \& {Brook}}{{Kawata} et~al.}{2017}]{Kawata17}
{Kawata} D.,  {Grand} R. J.~J.,  {Gibson} B.~K.,  {Casagrande} L.,  {Hunt} J.
  A.~S.,   {Brook} C.~B.,  2017, \mn@doi [\mnras] {10.1093/mnras/stw2363},
  \href {https://ui.adsabs.harvard.edu/abs/2017MNRAS.464..702K} {464, 702}

\bibitem[\protect\citeauthoryear{{Kawata}, {Baba}, {Ciuc{\v{a}}}, {Cropper},
  {Grand}, {Hunt}  \& {Seabroke}}{{Kawata} et~al.}{2018}]{Kawata18}
{Kawata} D.,  {Baba} J.,  {Ciuc{\v{a}}} I.,  {Cropper} M.,  {Grand} R. J.~J.,
  {Hunt} J. A.~S.,   {Seabroke} G.,  2018, \mn@doi [\mnras]
  {10.1093/mnrasl/sly107}, \href
  {https://ui.adsabs.harvard.edu/abs/2018MNRAS.479L.108K} {479, L108}

\bibitem[\protect\citeauthoryear{{Kordopatis} et~al.,}{{Kordopatis}
  et~al.}{2013}]{Kordopatis13}
{Kordopatis} G.,  et~al., 2013, \mn@doi [\mnras] {10.1093/mnras/stt1804}, \href
  {https://ui.adsabs.harvard.edu/abs/2013MNRAS.436.3231K} {436, 3231}

\bibitem[\protect\citeauthoryear{{Kordopatis} et~al.,}{{Kordopatis}
  et~al.}{2015}]{Kordopatis15}
{Kordopatis} G.,  et~al., 2015, \mn@doi [\mnras] {10.1093/mnras/stu2726}, \href
  {https://ui.adsabs.harvard.edu/abs/2015MNRAS.447.3526K} {447, 3526}

\bibitem[\protect\citeauthoryear{{Lee} et~al.,}{{Lee} et~al.}{2011}]{Lee11}
{Lee} Y.~S.,  et~al., 2011, \mn@doi [\apj] {10.1088/0004-637X/738/2/187}, \href
  {https://ui.adsabs.harvard.edu/abs/2011ApJ...738..187L} {738, 187}

\bibitem[\protect\citeauthoryear{{Li} \& {Shen}}{{Li} \& {Shen}}{2020}]{Li20}
{Li} Z.-Y.,  {Shen} J.,  2020, \mn@doi [\apj] {10.3847/1538-4357/ab6b21}, \href
  {https://ui.adsabs.harvard.edu/abs/2020ApJ...890...85L} {890, 85}

\bibitem[\protect\citeauthoryear{{Li} \& {Zhao}}{{Li} \& {Zhao}}{2017}]{Li17}
{Li} C.,  {Zhao} G.,  2017, \mn@doi [\apj] {10.3847/1538-4357/aa93f4}, \href
  {https://ui.adsabs.harvard.edu/abs/2017ApJ...850...25L} {850, 25}

\bibitem[\protect\citeauthoryear{{Lian} et~al.,}{{Lian} et~al.}{2020}]{Lian20}
{Lian} J.,  et~al., 2020, \mn@doi [\mnras] {10.1093/mnras/staa2205}, \href
  {https://ui.adsabs.harvard.edu/abs/2020MNRAS.497.3557L} {497, 3557}

\bibitem[\protect\citeauthoryear{{Lindegren} et~al.,}{{Lindegren}
  et~al.}{2018}]{Lindegren18}
{Lindegren} L.,  et~al., 2018, \mn@doi [\aap] {10.1051/0004-6361/201832727},
  \href {https://ui.adsabs.harvard.edu/abs/2018A&A...616A...2L} {616, A2}

\bibitem[\protect\citeauthoryear{{Liu}, {Xue}, {Fang}, {van de Ven}, {Wu},
  {Smith}  \& {Carrell}}{{Liu} et~al.}{2012}]{Liu12}
{Liu} C.,  {Xue} X.,  {Fang} M.,  {van de Ven} G.,  {Wu} Y.,  {Smith} M.~C.,
  {Carrell} K.,  2012, \mn@doi [\apjl] {10.1088/2041-8205/753/1/L24}, \href
  {https://ui.adsabs.harvard.edu/abs/2012ApJ...753L..24L} {753, L24}

\bibitem[\protect\citeauthoryear{{Liu} et~al.,}{{Liu} et~al.}{2014}]{Liu14}
{Liu} X.~W.,  et~al., 2014, in {Feltzing} S.,  {Zhao} G.,  {Walton} N.~A.,
  {Whitelock} P.,  eds,  IAU Symposium Vol. 298, Setting the scene for Gaia and
  LAMOST. pp 310--321 (\mn@eprint {arXiv} {1306.5376}),
  \mn@doi{10.1017/S1743921313006510}

\bibitem[\protect\citeauthoryear{{Loebman}, {Ro{\v{s}}kar}, {Debattista},
  {Ivezi{\'c}}, {Quinn}  \& {Wadsley}}{{Loebman} et~al.}{2011}]{Loebman11}
{Loebman} S.~R.,  {Ro{\v{s}}kar} R.,  {Debattista} V.~P.,  {Ivezi{\'c}}
  {\v{Z}}.,  {Quinn} T.~R.,   {Wadsley} J.,  2011, \mn@doi [\apj]
  {10.1088/0004-637X/737/1/8}, \href
  {https://ui.adsabs.harvard.edu/abs/2011ApJ...737....8L} {737, 8}

\bibitem[\protect\citeauthoryear{{Mackereth} et~al.,}{{Mackereth}
  et~al.}{2017}]{Mackereth17}
{Mackereth} J.~T.,  et~al., 2017, \mn@doi [\mnras] {10.1093/mnras/stx1774},
  \href {https://ui.adsabs.harvard.edu/abs/2017MNRAS.471.3057M} {471, 3057}

\bibitem[\protect\citeauthoryear{{Majewski} et~al.,}{{Majewski}
  et~al.}{2017}]{Majewski17}
{Majewski} S.~R.,  et~al., 2017, \mn@doi [\aj] {10.3847/1538-3881/aa784d},
  \href {https://ui.adsabs.harvard.edu/abs/2017AJ....154...94M} {154, 94}

\bibitem[\protect\citeauthoryear{{Minchev}, {Chiappini}  \& {Martig}}{{Minchev}
  et~al.}{2013}]{Minchev13}
{Minchev} I.,  {Chiappini} C.,   {Martig} M.,  2013, \mn@doi [\aap]
  {10.1051/0004-6361/201220189}, \href
  {https://ui.adsabs.harvard.edu/abs/2013A&A...558A...9M} {558, A9}

\bibitem[\protect\citeauthoryear{{Minchev}, {Martig}, {Streich}, {Scannapieco},
  {de Jong}  \& {Steinmetz}}{{Minchev} et~al.}{2015}]{Minchev15}
{Minchev} I.,  {Martig} M.,  {Streich} D.,  {Scannapieco} C.,  {de Jong} R.~S.,
    {Steinmetz} M.,  2015, \mn@doi [\apjl] {10.1088/2041-8205/804/1/L9}, \href
  {https://ui.adsabs.harvard.edu/abs/2015ApJ...804L...9M} {804, L9}

\bibitem[\protect\citeauthoryear{{Minchev}, {Chiappini}  \& {Martig}}{{Minchev}
  et~al.}{2016}]{Minchev16}
{Minchev} I.,  {Chiappini} C.,   {Martig} M.,  2016, \mn@doi [Astronomische
  Nachrichten] {10.1002/asna.201612404}, \href
  {https://ui.adsabs.harvard.edu/abs/2016AN....337..944M} {337, 944}

\bibitem[\protect\citeauthoryear{{Minchev} et~al.,}{{Minchev}
  et~al.}{2018}]{Minchev18}
{Minchev} I.,  et~al., 2018, \mn@doi [\mnras] {10.1093/mnras/sty2033}, \href
  {https://ui.adsabs.harvard.edu/abs/2018MNRAS.481.1645M} {481, 1645}

\bibitem[\protect\citeauthoryear{{Ness} et~al.,}{{Ness} et~al.}{2018}]{Ness18}
{Ness} M.,  et~al., 2018, \mn@doi [\apj] {10.3847/1538-4357/aa9d8e}, \href
  {https://ui.adsabs.harvard.edu/abs/2018ApJ...853..198N} {853, 198}

\bibitem[\protect\citeauthoryear{{Ness}, {Johnston}, {Blancato}, {Rix},
  {Beane}, {Bird}  \& {Hawkins}}{{Ness} et~al.}{2019}]{Ness19}
{Ness} M.~K.,  {Johnston} K.~V.,  {Blancato} K.,  {Rix} H.~W.,  {Beane} A.,
  {Bird} J.~C.,   {Hawkins} K.,  2019, \mn@doi [\apj]
  {10.3847/1538-4357/ab3e3c}, \href
  {https://ui.adsabs.harvard.edu/abs/2019ApJ...883..177N} {883, 177}

\bibitem[\protect\citeauthoryear{{Nordstr{\"o}m} et~al.,}{{Nordstr{\"o}m}
  et~al.}{2004}]{Nordstrom04}
{Nordstr{\"o}m} B.,  et~al., 2004, \mn@doi [\aap] {10.1051/0004-6361:20035959},
  \href {https://ui.adsabs.harvard.edu/abs/2004A&A...418..989N} {418, 989}

\bibitem[\protect\citeauthoryear{{Peng}, {Wu}, {Qi}, {Du}, {Ma}, {Zhou}, {Jia}
  \& {Wang}}{{Peng} et~al.}{2018}]{Peng18}
{Peng} X.,  {Wu} Z.,  {Qi} Z.,  {Du} C.,  {Ma} J.,  {Zhou} X.,  {Jia} Y.,
  {Wang} S.,  2018, \mn@doi [\pasp] {10.1088/1538-3873/aac1b5}, \href
  {https://ui.adsabs.harvard.edu/abs/2018PASP..130g4102P} {130, 074102}

\bibitem[\protect\citeauthoryear{{Poggio} et~al.,}{{Poggio}
  et~al.}{2018}]{Poggio18}
{Poggio} E.,  et~al., 2018, \mn@doi [\mnras] {10.1093/mnrasl/sly148}, \href
  {https://ui.adsabs.harvard.edu/abs/2018MNRAS.481L..21P} {481, L21}

\bibitem[\protect\citeauthoryear{{Quillen}, {Nolting}, {Minchev}, {De Silva}
  \& {Chiappini}}{{Quillen} et~al.}{2018}]{Quillen18}
{Quillen} A.~C.,  {Nolting} E.,  {Minchev} I.,  {De Silva} G.,   {Chiappini}
  C.,  2018, \mn@doi [\mnras] {10.1093/mnras/sty125}, \href
  {https://ui.adsabs.harvard.edu/abs/2018MNRAS.475.4450Q} {475, 4450}

\bibitem[\protect\citeauthoryear{{Recio-Blanco} et~al.,}{{Recio-Blanco}
  et~al.}{2014}]{Recio14}
{Recio-Blanco} A.,  et~al., 2014, \mn@doi [\aap] {10.1051/0004-6361/201322944},
  \href {https://ui.adsabs.harvard.edu/abs/2014A&A...567A...5R} {567, A5}

\bibitem[\protect\citeauthoryear{{Romero-G{\'o}mez}, {Mateu}, {Aguilar},
  {Figueras}  \& {Castro-Ginard}}{{Romero-G{\'o}mez} et~al.}{2019}]{Romero19}
{Romero-G{\'o}mez} M.,  {Mateu} C.,  {Aguilar} L.,  {Figueras} F.,
  {Castro-Ginard} A.,  2019, \mn@doi [\aap] {10.1051/0004-6361/201834908},
  \href {https://ui.adsabs.harvard.edu/abs/2019A&A...627A.150R} {627, A150}

\bibitem[\protect\citeauthoryear{{Ro{\v{s}}kar}, {Debattista}, {Quinn},
  {Stinson}  \& {Wadsley}}{{Ro{\v{s}}kar} et~al.}{2008}]{Roskar08}
{Ro{\v{s}}kar} R.,  {Debattista} V.~P.,  {Quinn} T.~R.,  {Stinson} G.~S.,
  {Wadsley} J.,  2008, \mn@doi [\apjl] {10.1086/592231}, \href
  {https://ui.adsabs.harvard.edu/abs/2008ApJ...684L..79R} {684, L79}

\bibitem[\protect\citeauthoryear{{Sch{\"o}nrich} \& {Binney}}{{Sch{\"o}nrich}
  \& {Binney}}{2009}]{Schonrich09}
{Sch{\"o}nrich} R.,  {Binney} J.,  2009, \mn@doi [\mnras]
  {10.1111/j.1365-2966.2009.14750.x}, \href
  {https://ui.adsabs.harvard.edu/abs/2009MNRAS.396..203S} {396, 203}

\bibitem[\protect\citeauthoryear{{Sellwood} \& {Binney}}{{Sellwood} \&
  {Binney}}{2002}]{Sellwood02}
{Sellwood} J.~A.,  {Binney} J.~J.,  2002, \mn@doi [\mnras]
  {10.1046/j.1365-8711.2002.05806.x}, \href
  {https://ui.adsabs.harvard.edu/abs/2002MNRAS.336..785S} {336, 785}

\bibitem[\protect\citeauthoryear{{Siebert} et~al.,}{{Siebert}
  et~al.}{2011}]{Siebert11}
{Siebert} A.,  et~al., 2011, \mn@doi [\mnras]
  {10.1111/j.1365-2966.2010.18037.x}, \href
  {https://ui.adsabs.harvard.edu/abs/2011MNRAS.412.2026S} {412, 2026}

\bibitem[\protect\citeauthoryear{{Skowron} et~al.,}{{Skowron}
  et~al.}{2019}]{Skowron19}
{Skowron} D.~M.,  et~al., 2019, \mn@doi [\actaa] {10.32023/0001-5237/69.4.1},
  \href {https://ui.adsabs.harvard.edu/abs/2019AcA....69..305S} {69, 305}

\bibitem[\protect\citeauthoryear{{Spagna}, {Lattanzi}, {Re Fiorentin}  \&
  {Smart}}{{Spagna} et~al.}{2010}]{Spagna10}
{Spagna} A.,  {Lattanzi} M.~G.,  {Re Fiorentin} P.,   {Smart} R.~L.,  2010,
  \mn@doi [\aap] {10.1051/0004-6361/200913538}, \href
  {https://ui.adsabs.harvard.edu/abs/2010A&A...510L...4S} {510, L4}

\bibitem[\protect\citeauthoryear{{Spitoni}, {Silva Aguirre}, {Matteucci},
  {Calura}  \& {Grisoni}}{{Spitoni} et~al.}{2019}]{Spitoni19}
{Spitoni} E.,  {Silva Aguirre} V.,  {Matteucci} F.,  {Calura} F.,   {Grisoni}
  V.,  2019, \mn@doi [\aap] {10.1051/0004-6361/201834188}, \href
  {https://ui.adsabs.harvard.edu/abs/2019A&A...623A..60S} {623, A60}

\bibitem[\protect\citeauthoryear{{Steinmetz} et~al.,}{{Steinmetz}
  et~al.}{2006}]{Steinmetz06}
{Steinmetz} M.,  et~al., 2006, \mn@doi [\aj] {10.1086/506564}, \href
  {https://ui.adsabs.harvard.edu/abs/2006AJ....132.1645S} {132, 1645}

\bibitem[\protect\citeauthoryear{{Sun} et~al.,}{{Sun} et~al.}{2015}]{Sun15}
{Sun} N.-C.,  et~al., 2015, \mn@doi [Research in Astronomy and Astrophysics]
  {10.1088/1674-4527/15/8/017}, \href
  {https://ui.adsabs.harvard.edu/abs/2015RAA....15.1342S} {15, 1342}

\bibitem[\protect\citeauthoryear{{Tian} et~al.,}{{Tian} et~al.}{2015}]{Tian15}
{Tian} H.-J.,  et~al., 2015, \mn@doi [\apj] {10.1088/0004-637X/809/2/145},
  \href {https://ui.adsabs.harvard.edu/abs/2015ApJ...809..145T} {809, 145}

\bibitem[\protect\citeauthoryear{{Tian} et~al.,}{{Tian} et~al.}{2017}]{Tian17}
{Tian} H.-J.,  et~al., 2017, \mn@doi [Research in Astronomy and Astrophysics]
  {10.1088/1674-4527/17/11/114}, \href
  {https://ui.adsabs.harvard.edu/abs/2017RAA....17..114T} {17, 114}

\bibitem[\protect\citeauthoryear{{Tian}, {Liu}, {Wu}, {Xiang}  \&
  {Zhang}}{{Tian} et~al.}{2018}]{Tian18}
{Tian} H.-J.,  {Liu} C.,  {Wu} Y.,  {Xiang} M.-S.,   {Zhang} Y.,  2018, \mn@doi
  [\apjl] {10.3847/2041-8213/aae1f3}, \href
  {https://ui.adsabs.harvard.edu/abs/2018ApJ...865L..19T} {865, L19}

\bibitem[\protect\citeauthoryear{{Ting} \& {Rix}}{{Ting} \&
  {Rix}}{2019}]{Ting19}
{Ting} Y.-S.,  {Rix} H.-W.,  2019, \mn@doi [\apj] {10.3847/1538-4357/ab1ea5},
  \href {https://ui.adsabs.harvard.edu/abs/2019ApJ...878...21T} {878, 21}

\bibitem[\protect\citeauthoryear{{Vrard}, {Mosser}  \& {Samadi}}{{Vrard}
  et~al.}{2016}]{Vrard16}
{Vrard} M.,  {Mosser} B.,   {Samadi} R.,  2016, \mn@doi [\aap]
  {10.1051/0004-6361/201527259}, \href
  {https://ui.adsabs.harvard.edu/abs/2016A&A...588A..87V} {588, A87}

\bibitem[\protect\citeauthoryear{{Wang}, {L{\'o}pez-Corredoira}, {Carlin}  \&
  {Deng}}{{Wang} et~al.}{2018a}]{Wang18b}
{Wang} H.,  {L{\'o}pez-Corredoira} M.,  {Carlin} J.~L.,   {Deng} L.,  2018a,
  \mn@doi [\mnras] {10.1093/mnras/sty739}, \href
  {https://ui.adsabs.harvard.edu/abs/2018MNRAS.477.2858W} {477, 2858}

\bibitem[\protect\citeauthoryear{{Wang}, {Liu}, {Xu}, {Wan}  \& {Deng}}{{Wang}
  et~al.}{2018b}]{Wang18a}
{Wang} H.-F.,  {Liu} C.,  {Xu} Y.,  {Wan} J.-C.,   {Deng} L.,  2018b, \mn@doi
  [\mnras] {10.1093/mnras/sty1058}, \href
  {https://ui.adsabs.harvard.edu/abs/2018MNRAS.478.3367W} {478, 3367}

\bibitem[\protect\citeauthoryear{{Wang} et~al.,}{{Wang} et~al.}{2019a}]{Wang19}
{Wang} C.,  et~al., 2019a, \mn@doi [\apjl] {10.3847/2041-8213/ab1fdd}, \href
  {https://ui.adsabs.harvard.edu/abs/2019ApJ...877L...7W} {877, L7}

\bibitem[\protect\citeauthoryear{{Wang} et~al.,}{{Wang} et~al.}{2019b}]{Wh19}
{Wang} H.-F.,  et~al., 2019b, \mn@doi [\apj] {10.3847/1538-4357/ab4204}, \href
  {https://ui.adsabs.harvard.edu/abs/2019ApJ...884..135W} {884, 135}

\bibitem[\protect\citeauthoryear{{Wang} et~al.,}{{Wang} et~al.}{2020}]{Wang20}
{Wang} H.~F.,  et~al., 2020, \mn@doi [\apj] {10.3847/1538-4357/ab93ad}, \href
  {https://ui.adsabs.harvard.edu/abs/2020ApJ...897..119W} {897, 119}

\bibitem[\protect\citeauthoryear{{Widrow}, {Gardner}, {Yanny}, {Dodelson}  \&
  {Chen}}{{Widrow} et~al.}{2012}]{Widrow12}
{Widrow} L.~M.,  {Gardner} S.,  {Yanny} B.,  {Dodelson} S.,   {Chen} H.-Y.,
  2012, \mn@doi [\apjl] {10.1088/2041-8205/750/2/L41}, \href
  {https://ui.adsabs.harvard.edu/abs/2012ApJ...750L..41W} {750, L41}

\bibitem[\protect\citeauthoryear{{Widrow}, {Barber}, {Chequers}  \&
  {Cheng}}{{Widrow} et~al.}{2014}]{Widrow14}
{Widrow} L.~M.,  {Barber} J.,  {Chequers} M.~H.,   {Cheng} E.,  2014, \mn@doi
  [\mnras] {10.1093/mnras/stu396}, \href
  {https://ui.adsabs.harvard.edu/abs/2014MNRAS.440.1971W} {440, 1971}

\bibitem[\protect\citeauthoryear{{Williams} et~al.,}{{Williams}
  et~al.}{2013}]{Williams13}
{Williams} M.~E.~K.,  et~al., 2013, \mn@doi [\mnras] {10.1093/mnras/stt1522},
  \href {https://ui.adsabs.harvard.edu/abs/2013MNRAS.436..101W} {436, 101}

\bibitem[\protect\citeauthoryear{{Wojno} et~al.,}{{Wojno}
  et~al.}{2018}]{Wojno18}
{Wojno} J.,  et~al., 2018, \mn@doi [\mnras] {10.1093/mnras/sty1016}, \href
  {https://ui.adsabs.harvard.edu/abs/2018MNRAS.477.5612W} {477, 5612}

\bibitem[\protect\citeauthoryear{{Wu} et~al.,}{{Wu} et~al.}{2018}]{Wu18}
{Wu} Y.,  et~al., 2018, \mn@doi [\mnras] {10.1093/mnras/stx3296}, \href
  {https://ui.adsabs.harvard.edu/abs/2018MNRAS.475.3633W} {475, 3633}

\bibitem[\protect\citeauthoryear{{Wu} et~al.,}{{Wu} et~al.}{2019}]{Wu19}
{Wu} Y.,  et~al., 2019, \mn@doi [\mnras] {10.1093/mnras/stz256}, \href
  {https://ui.adsabs.harvard.edu/abs/2019MNRAS.484.5315W} {484, 5315}

\bibitem[\protect\citeauthoryear{{Xiang} et~al.,}{{Xiang}
  et~al.}{2015a}]{Xiang15b}
{Xiang} M.-S.,  et~al., 2015a, \mn@doi [Research in Astronomy and Astrophysics]
  {10.1088/1674-4527/15/8/009}, \href
  {https://ui.adsabs.harvard.edu/abs/2015RAA....15.1209X} {15, 1209}

\bibitem[\protect\citeauthoryear{{Xiang} et~al.,}{{Xiang}
  et~al.}{2015b}]{Xiang15a}
{Xiang} M.~S.,  et~al., 2015b, \mn@doi [\mnras] {10.1093/mnras/stu2692}, \href
  {https://ui.adsabs.harvard.edu/abs/2015MNRAS.448..822X} {448, 822}

\bibitem[\protect\citeauthoryear{{Xiang} et~al.,}{{Xiang}
  et~al.}{2017a}]{Xiang17a}
{Xiang} M.,  et~al., 2017a, \mn@doi [\apjs] {10.3847/1538-4365/aa80e4}, \href
  {https://ui.adsabs.harvard.edu/abs/2017ApJS..232....2X} {232, 2}

\bibitem[\protect\citeauthoryear{{Xiang} et~al.,}{{Xiang}
  et~al.}{2017b}]{Xiang17b}
{Xiang} M.~S.,  et~al., 2017b, \mn@doi [\mnras] {10.1093/mnras/stx129}, \href
  {https://ui.adsabs.harvard.edu/abs/2017MNRAS.467.1890X} {467, 1890}

\bibitem[\protect\citeauthoryear{{Xiang} et~al.,}{{Xiang}
  et~al.}{2018}]{Xiang18}
{Xiang} M.,  et~al., 2018, \mn@doi [\apjs] {10.3847/1538-4365/aad237}, \href
  {https://ui.adsabs.harvard.edu/abs/2018ApJS..237...33X} {237, 33}

\bibitem[\protect\citeauthoryear{{Xu}, {Newberg}, {Carlin}, {Liu}, {Deng},
  {Li}, {Sch{\"o}nrich}  \& {Yanny}}{{Xu} et~al.}{2015}]{Xu15}
{Xu} Y.,  {Newberg} H.~J.,  {Carlin} J.~L.,  {Liu} C.,  {Deng} L.,  {Li} J.,
  {Sch{\"o}nrich} R.,   {Yanny} B.,  2015, \mn@doi [\apj]
  {10.1088/0004-637X/801/2/105}, \href
  {https://ui.adsabs.harvard.edu/abs/2015ApJ...801..105X} {801, 105}

\bibitem[\protect\citeauthoryear{{Yan}, {Du}, {Liu}, {Li}, {Shi}, {Chen}, {Ma}
  \& {Wu}}{{Yan} et~al.}{2019}]{Yan19}
{Yan} Y.,  {Du} C.,  {Liu} S.,  {Li} H.,  {Shi} J.,  {Chen} Y.,  {Ma} J.,
  {Wu} Z.,  2019, \mn@doi [\apj] {10.3847/1538-4357/ab287d}, \href
  {https://ui.adsabs.harvard.edu/abs/2019ApJ...880...36Y} {880, 36}

\bibitem[\protect\citeauthoryear{{Yanny} et~al.,}{{Yanny}
  et~al.}{2009}]{Yanny09}
{Yanny} B.,  et~al., 2009, \mn@doi [\apj] {10.1088/0004-637X/700/2/1282}, \href
  {https://ui.adsabs.harvard.edu/abs/2009ApJ...700.1282Y} {700, 1282}

\bibitem[\protect\citeauthoryear{{Yuan} et~al.,}{{Yuan} et~al.}{2015}]{Yuan15}
{Yuan} H.~B.,  et~al., 2015, \mn@doi [\mnras] {10.1093/mnras/stu2723}, \href
  {https://ui.adsabs.harvard.edu/abs/2015MNRAS.448..855Y} {448, 855}

\bibitem[\protect\citeauthoryear{{Zhao}, {Chen}, {Shi}, {Liang}, {Hou}, {Chen},
  {Zhang}  \& {Li}}{{Zhao} et~al.}{2006}]{Zhao06}
{Zhao} G.,  {Chen} Y.-Q.,  {Shi} J.-R.,  {Liang} Y.-C.,  {Hou} J.-L.,  {Chen}
  L.,  {Zhang} H.-W.,   {Li} A.-G.,  2006, \mn@doi [\cjaa]
  {10.1088/1009-9271/6/3/01}, \href
  {https://ui.adsabs.harvard.edu/abs/2006ChJAA...6..265Z} {6, 265}

\bibitem[\protect\citeauthoryear{{Zhao}, {Zhao}, {Chu}, {Jing}  \&
  {Deng}}{{Zhao} et~al.}{2012}]{Zhao12}
{Zhao} G.,  {Zhao} Y.-H.,  {Chu} Y.-Q.,  {Jing} Y.-P.,   {Deng} L.-C.,  2012,
  \mn@doi [Research in Astronomy and Astrophysics]
  {10.1088/1674-4527/12/7/002}, \href
  {https://ui.adsabs.harvard.edu/abs/2012RAA....12..723Z} {12, 723}

\makeatother
\end{thebibliography}

\appendix
\section{}

Fig\,A1 shows an example of the 1D distribution of $V_{R}$, $J_{R}$, $V_{\phi}$, $J_{\phi}$, $V_{Z}$ and $J_{Z}$ for different age and metallicity bins. In particular, we select two age and metallicity bins, and mark the value of skewness of these distribution in the figure.
\begin{figure*}
\centering
\includegraphics[width=\linewidth]{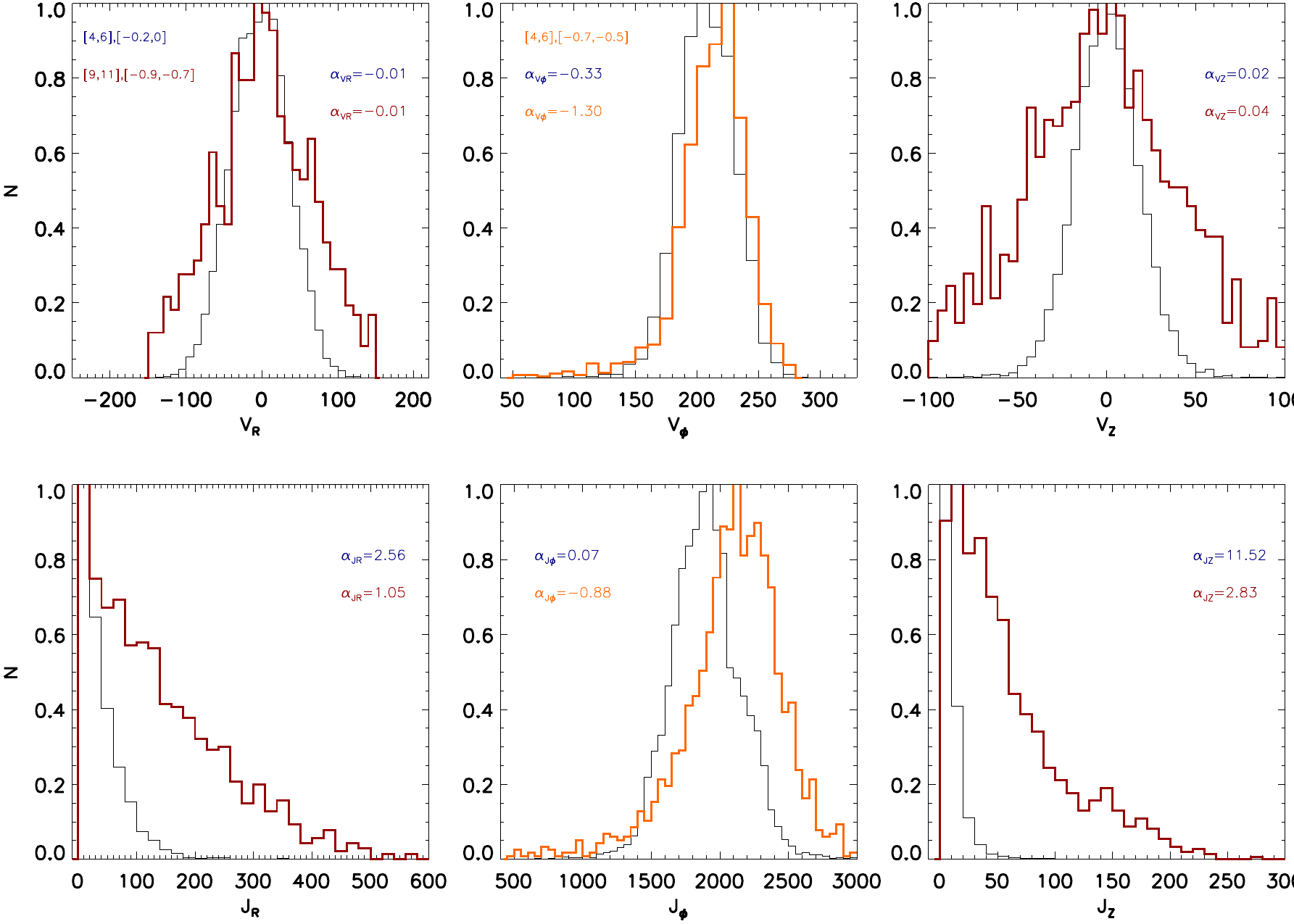}
\caption{1D distribution of $V_{R}$, $J_{R}$, $V_{\phi}$, $J_{\phi}$, $V_{Z}$ and $J_{Z}$ for different age bins and metallicity bins, black line represents 4 $<$ $\tau$ $<$ 6\,Gyr, $-$0.2 $<$ [Fe/H] $<$ 0, red line represents 9 $<$ $\tau$ $<$ 11\,Gyr, $-$0.9 $<$ [Fe/H] $<$ $-$0.7\,dex. Distributions with positive tails exhibit positive values of skewness, while distributions with negative tails exhibit negative values of skewness.  \label{fig:figA}}
\end{figure*}

Fig.\,A2 shows the distributions at different birth radius in the $R$ -- $Z$ plane. We select three regions in birth radii to validate the median distance alteration due to the radial migration shown in Fig.\,15. From Fig.\,A2, we can see that for stars born beyond 10\,kpc, the results are consistent with Fig.\,15, whereas stars in the outer disc show much stronger inward migration effect. This may be caused by the stronger interactions with satellite galaxies at the outer disc.
\begin{figure*}
\centering
\includegraphics[width=\linewidth]{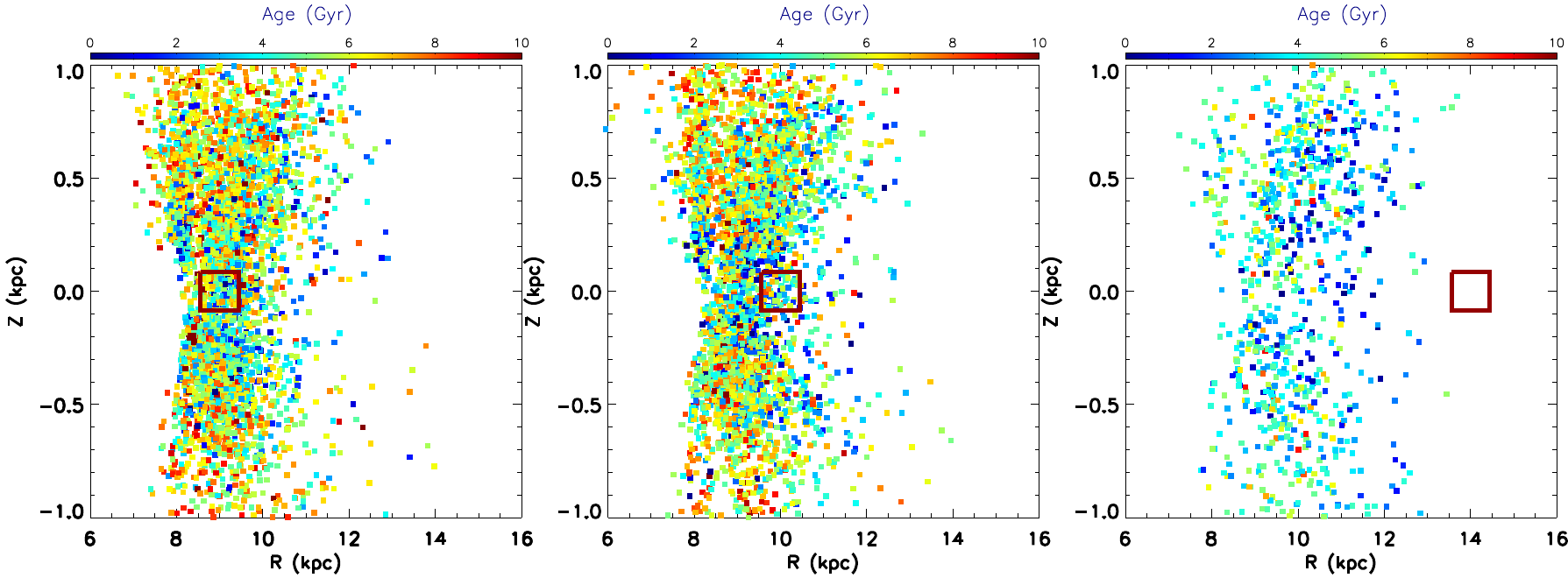}
\caption{Distributions in the $R$ -- $Z$ plane for stars born at 9\,kpc,({\em left}), 10\,kpc ({\em middle}), 14\,kpc ({\em right}), as indicated by the red box. The color represents stellar age. \label{fig:fig1}}
\end{figure*}

\label{lastpage}

\end{document}